\documentclass[showpacs,showkeys,amsmath,amssymb,twocolumn,pra,superscriptaddress,10pt]{revtex4-2}

\usepackage{amsfonts,amsthm}
\usepackage{subfigure}
\usepackage{amssymb}
\usepackage{amscd}
\usepackage{amsmath}  
\usepackage{multirow}
\usepackage{graphicx}
\usepackage{dcolumn}
\usepackage{bm}
\usepackage{bbm}
\usepackage{cleveref}

\newcommand{\bra}[1]{\mbox{$\left\langle #1 \right|$}}
\newcommand{\ket}[1]{\mbox{$\left| #1 \right\rangle$}}

\usepackage{color}

\begin{document}
\title{Deep Ising Born Machine}
\author{Zhu Cao}
\email{caozhu@ecust.edu.cn}
\address{Key Laboratory of Smart Manufacturing in Energy Chemical Process, Ministry of Education, East China University of Science and Technology, Shanghai 200237, China}

\begin{abstract}
A quantum neural network (QNN) is a method to find patterns in quantum data and has a wide range of applications including quantum chemistry, quantum computation, quantum metrology, and quantum simulation.  Efficiency and universality are two desirable properties of a QNN but are unfortunately contradictory. In this work, we examine a \emph{deep Ising Born machine} (DIBoM), and show it has a good balance between efficiency and universality. More precisely, the DIBoM has a flexible number of parameters to be efficient, and achieves provable universality with sufficient parameters. The architecture of the DIBoM is based on generalized controlled-Z gates, conditional gates,  and some other ingredients. To compare the universality of the DIBoM with other QNNs, we propose a fidelity-based expressivity measure, which may be of independent interest. Extensive empirical evaluations corroborate that the DIBoM is both efficient and expressive.

\end{abstract}



\vspace{2pc}

\maketitle

\section{Introduction}

Machine learning (ML) has emerged as one of the most revolutionary techniques in recent years \cite{Goodfellow-et-al-2016}. Despite its significance, ML necessitates a tremendous amount of computational power. However, with the waning effectiveness of Moore's law on the speed of classical processors \cite{waldrop2016chips}, and the ever-increasing computational demands of state-of-the-art ML models, the future development of ML may face significant hindrances due to the shortage of adequate computational resources. Quantum computing \cite{nielsen2010quantum}, a novel computing paradigm, holds the potential to sustainably advance ML. At present, quantum machine learning (QML) \cite{biamonte2017quantum}, which refers to the use of quantum computers for machine learning, is still in its nascent stage. Depending on whether the data and the learning algorithm are classical or quantum, QML can be categorized into four types: classical learning of classical data, quantum learning of classical data, classical learning of quantum data, and quantum learning of quantum data.

Among the four categories of QML, quantum learning of quantum data is arguably the most promising type to achieve a demonstrable exponential speedup over classical machine learning methods. Furthermore, quantum learning of quantum data has a diverse array of applications, including quantum chemistry \cite{arute2020hartree}, quantum data compression \cite{romero2017quantum}, quantum error correction \cite{johnson2017qvector}, quantum metrology \cite{koczor2020variational}, quantum compiling \cite{sharma2020noise}, quantum state diagonalization \cite{larose2019variational}, quantum simulation \cite{li2017efficient}, quantum fidelity estimation \cite{cerezo2020variational}, and consistent histories \cite{arrasmith2019variational}. It is worth noting that these quantum applications generate a substantial amount of quantum data, which in turn fuels the development of quantum learning of quantum data, analogous to how the vast amount of classical information drives the advancement of classical machine learning.

Methods of quantum learning of quantum data can be classified into two categories: those that belong to quantum neural networks (QNNs) and those that do not. Examples of methods that do not fall into the QNN category are the Harrow-Hassidim-Lloyd algorithm \cite{harrow2009quantum}, quantum principal component analysis \cite{lloyd2014quantum}, and quantum support vector machines \cite{rebentrost2014quantum}. Various proposals of QNNs have been put forward in the literature \cite{schuld2015simulating,lewenstein1994quantum,wan2017quantum,da2016quantum,gonsalves2017quantum,kouda2005qubit,beer2020training}, among which the current state-of-the-art is arguably given by Ref.~\cite{beer2020training}. In addition, specialized QNNs tailored to specific data inputs have also been developed, including quantum convolutional neural networks \cite{cong2019quantum}, quantum recurrent neural networks \cite{bausch2020recurrent}, quantum generative adversarial networks \cite{lloyd2018quantum}, quantum autoencoders \cite{bondarenko2020quantum}, quantum reservoir networks \cite{ghosh2021quantum}, and quantum residual networks \cite{killoran2019continuous}. However, these specialized QNNs cannot perform universal quantum computation, which is essential for the general quantum learning task of learning the hidden mapping between a set of quantum input and label pairs. Consequently, we will focus our attention on general QNNs and the general quantum learning task hereafter.

Efficiency and universality are two desirable properties of a general QNN. While efficiency is measured in terms of the number of parameters in the model, which should be as small as possible, universality refers to the ability of a QNN to approximate an arbitrary unitary on $n$ qubits. These two properties are often in conflict with each other. For instance, consider a basic QNN that applies a parametrized unitary $U$ to the quantum input and produces an output that approximates the label. Here, the parametrized unitary $U$ for $n$ qubits is represented as 
\begin{equation}
U = \exp \left[ i\left( \sum\limits_{j_1=0}^{3}\cdots\sum\limits_{j_n=0}^{3} \alpha_{j_1,j_2,\dots,j_n} \left( \sigma_{j_1}\otimes \cdots \otimes \sigma_{j_n} \right) \right) \right],
\label{eq:generalQNN}
\end{equation}
where $\sigma_0$ is the identity matrix, $\sigma_1$, $\sigma_2$, $\sigma_3$ are Pauli matrices, and $\alpha_{j_1,j_2,\dots,j_n}$ are real parameters that are learned during training. Although this basic QNN is universal, as it can approximate any unitary by adjusting its parameters, it is not efficient since its number of parameters is $4^n$. Ideally, the number of parameters should be polynomial in $n$ for the model to be efficient.

This work investigates a \emph{deep Ising Born machine} (DIBoM), which has a flexible number of parameters to mitigate the efficiency issue while retaining universality with sufficient parameters. The DIBoM consists of a generalized controlled-Z (CZ) gate, a conditional gate, a global or local cost function, and some other ingredients. By replacing the normal CZ gate with a generalized CZ gate in a hardware-efficient QNN \cite{mcclean2018barren,benedetti2019generative}, we demonstrate that the expressivity can be increased through numerical evaluations. Along the way, we develop an expressivity measure, called fidelity-based expressivity, to characterize the expressivity of different QNN architectures. This measure may be of independent interest. Moreover, we theoretically prove that hardware-efficient QNN with generalized CZ gates can achieve universal quantum computation with sufficient parameters. The conditional gate is used to solve the problem of different input and output dimensions, and this approach can save space by a constant factor compared to dissipative QNNs \cite{beer2020training}. In addition, the ablation study shows that this ingredient improves the expressivity of the DIBoM. We examine two variants of the DIBoM, one with a global cost function and the other with a local cost function, and show that the global cost function version has a wider range of applicability, while the local cost function is more trainable and can mitigate the barren plateau issue.
We perform extensive experiments to compare the DIBoM with other QNN architectures, evaluate its different components,  analyze the sensitivity of its performance to various parameters, and examine its robustness to noise. Our work invites further research on the design of QNNs with multiple desirable properties, and we hope it will stimulate further development of the architecture design of QNNs in general.

The roadmap for the rest of the paper is as follows. First, in Sec.~\ref{sec:relatedwork}, we review related works. Next, in  Sec.~\ref{sec:model}, we present the DIBoM model and its training method. In Sec.~\ref{sec:theory}, we analyze theoretically the properties of the DIBoM. We then turn to the empirical evaluation of the model, with Sec. \ref{sec:simulationsetup} describing the simulation setup and Sec. \ref{sec:simulationresult} presenting the results. 
Finally, we conclude the paper in Sec. \ref{sec:discussion} and give a few outlooks.

\section{Related works}
\label{sec:relatedwork}

In this section, we provide a review of the related works in the field, including hardware-efficient QNNs, dissipative QNNs, Ising Born machines, and Hamiltonian learning.

\subsection{Hardware-efficient QNNs}
\begin{figure}[htb]
\centering \includegraphics[width=8cm]{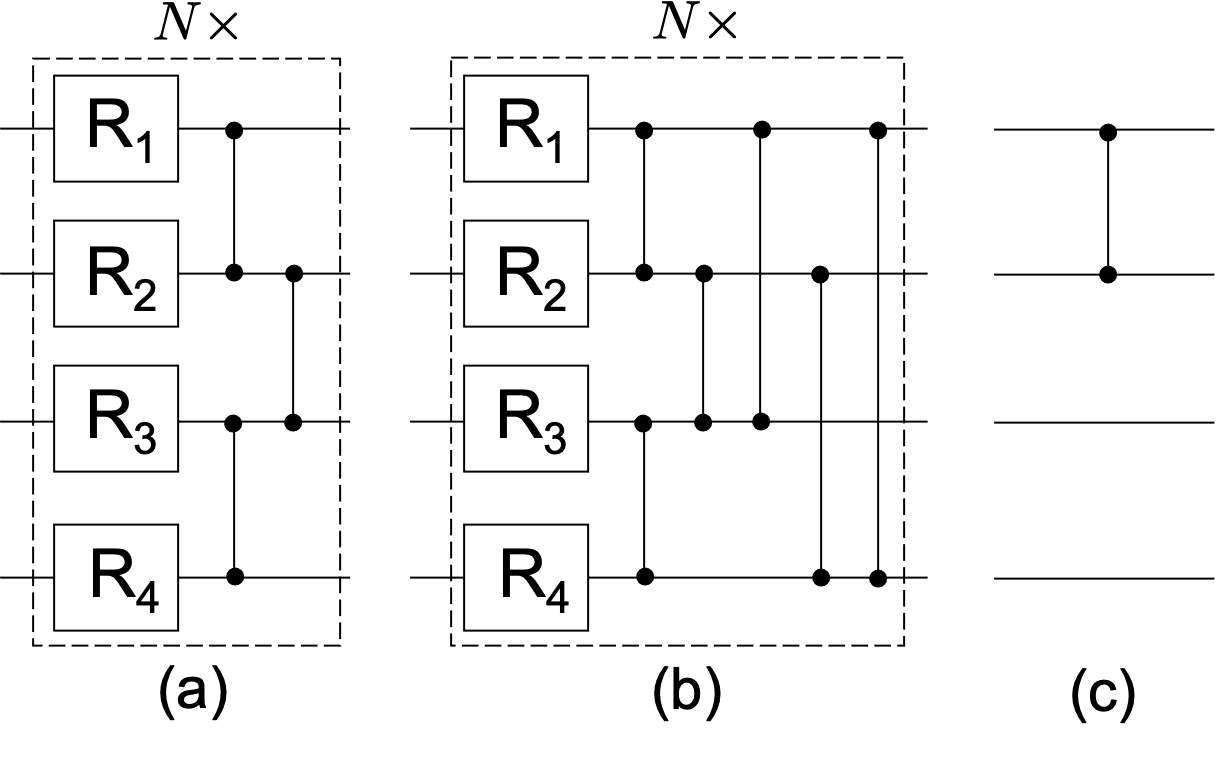}
\caption{(a) A series of blocks where each block consists of single-qubit rotations with nearest-neighbor CZ gates; (b) A series of blocks where each block consists of single-qubit rotations with all-to-all CZ gates; (c) Unitary with a single CZ gate connecting the first two qubits. }
\label{fig:illlus}
\end{figure}

We start by reviewing hardware-efficient QNNs \cite{mcclean2018barren,benedetti2019generative}. Hardware-efficient QNNs were proposed to reduce the exponential training cost of the basic QNN, and require only a polynomial number of resources. They are composed of alternating layers of single-qubit rotations and entangling gates such as CZ gates. 
The connectivity of the entangling gates can be either linear \cite{mcclean2018barren} or pairwise \cite{benedetti2019generative},  as shown in Figs.~\ref{fig:illlus}(a) and (b). The layer number of a hardware-efficient QNN can vary and so are the parameters of its single-qubit rotations. From now on, we will refer to the architecture in Ref.~\cite{mcclean2018barren} as \emph{the} hardware-efficient QNN.

There are two drawbacks to the hardware-efficient QNN. First, to our knowledge, there is to date no proof that the hardware-efficient QNN presented in \cite{mcclean2018barren} is capable of universal quantum computation even with an exponential number of layers, see Sec.~\ref{sec:power} for more details. In particular, it is not known whether it can be used to simulate a circuit with a single CZ gate, which is illustrated in Fig.~\ref{fig:illlus}(c). Secondly, it falls short of varying the relative number of qubits between the input and the output. The DIBoM resolves these shortcomings of the hardware-efficient QNN \cite{mcclean2018barren} while retaining its merits.

Note that the definition of the hardware-efficient QNN varies in the literature, and in the broadest sense can include any QNN that can be implemented efficiently on some quantum hardware. This in particular includes the DIBoM as a special case, which is different from its usage in our work.

\subsection{Dissipative QNN}
\begin{figure}[htb]
\centering \includegraphics[width=8cm]{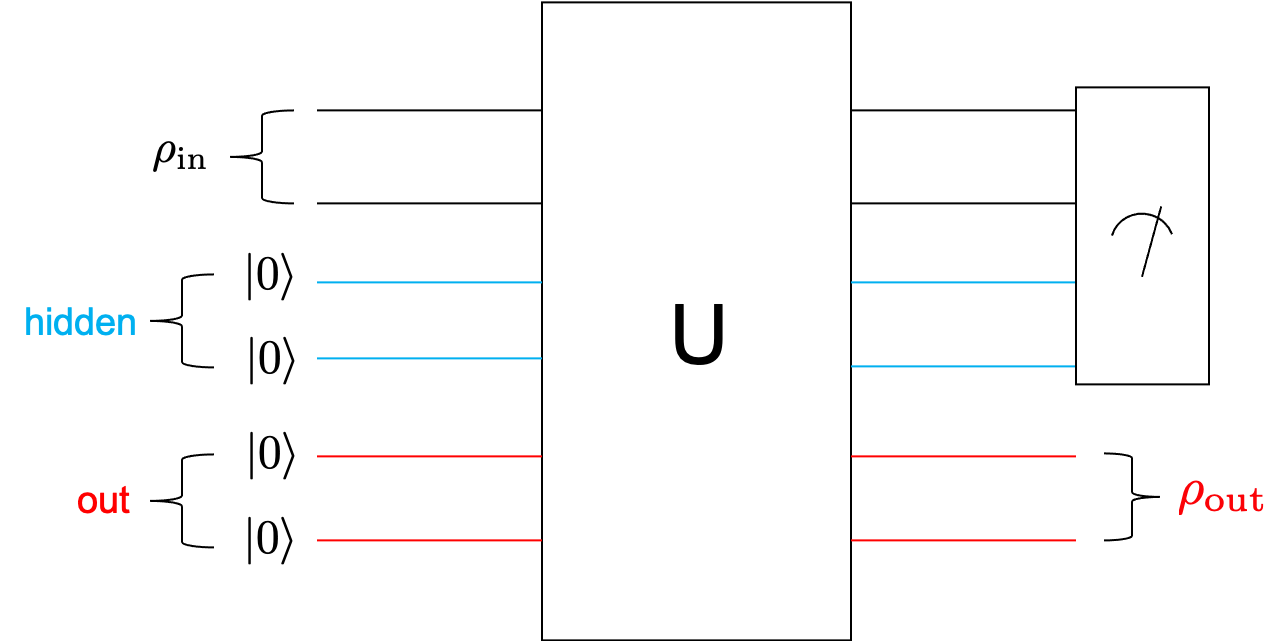}
\caption{An illustration of a dissipative QNN. Here, $\rho_{in}$ and $\rho_{out}$ are its quantum input and quantum output respectively, and $U$ is a unitary. The hidden and output qubits are all in the quantum state $\ket{0}$ initially.  }
\label{fig:dissipativeQNN}
\end{figure}

We next review dissipative QNNs, a different type of QNNs that deal with quantum inputs and outputs of unmatched dimensions \cite{beer2020training,sharma2020trainability}. Initially, all hidden and output qubits are in the state $\ket{0}$. Dissipative QNNs apply a unitary on all input, hidden, and output qubits  and subsequently trace out the input and hidden qubits, 
\begin{equation}
\rho_{out} \equiv \mathrm{tr}_{in,hid}\left( U( \rho_{in} \otimes \ket{0\cdots 0}_{hid,out} \bra{0\cdots 0} ) U^\dagger \right).
\end{equation}
This process is illustrated in Fig.~\ref{fig:dissipativeQNN}. Here, it can be easily seen that the dimensions of the quantum input $\rho_{in}$ and the quantum output $\rho_{out}$ need not be the same. Since this network architecture discards lots of qubits, it was given the name a \emph{dissipative QNN} \cite{sharma2020trainability}. 

However, a dissipative QNN also has several disadvantages. First, it has a larger space complexity due to its dissipative nature. When the dimensions of the quantum input and output are equal, a dissipative QNN has an overhead of 2 over a basic QNN in terms of the qubits used. Second, the resource that a dissipative QNN requires is still exponential, due to the same reason that an exponential number of parameters are needed to parametrize a unitary transformation. A DIBoM maintains the merit of a dissipative QNN that can handle unequal quantum input and output dimensions and in the meantime has the additional merits of having small space complexity and being resource efficient.

\subsection{Ising Born machine}
\begin{figure}[htb]
\centering \includegraphics[width=8.5cm]{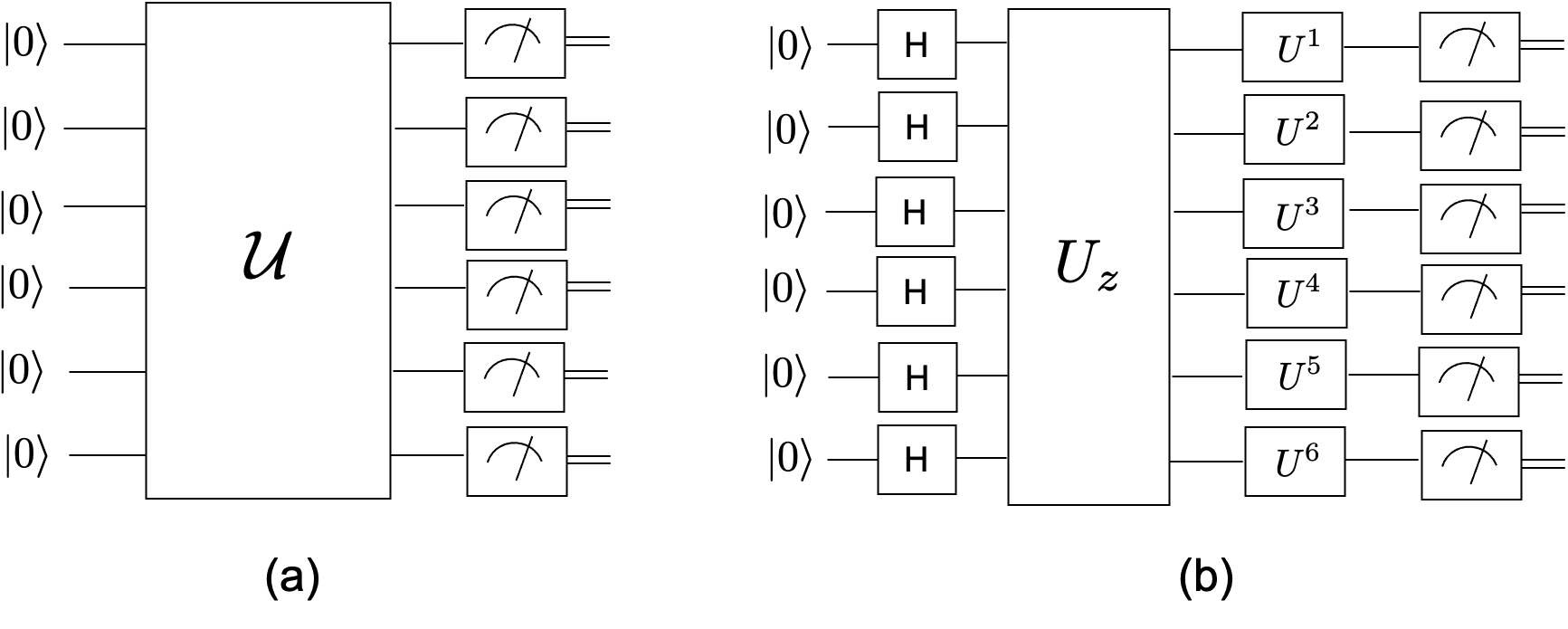}
\caption{(a) The schematics of a quantum Born machine. Here, $\mathcal{U}$ is a parametrized quantum circuit. (b) The schematics of an Ising Born machine. Here, $H$ is a Hadamard gate, $U_z$ consists of tunable two-qubit gates on all pairs of qubits, and $U^k$ ($k\in \mathbb{N}$) is a tunable single-qubit gate on qubit $k$. }
\label{fig:isingborn}
\end{figure}

We then review the Ising Born machine \cite{coyle2020born}, which bears a close resemblance to the DIBoM. We begin by defining a quantum Born machine \cite{cheng2018information,liu2018differentiable,benedetti2019generative}, as illustrated in Fig.~\ref{fig:isingborn}(a). A quantum Born machine is a class of models that consists of a parametrized quantum circuit followed by a quantum measurement. As the measurement outcome is determined by Born's rule, these models are named ``quantum Born machines'' \cite{cheng2018information,liu2018differentiable,benedetti2019generative}. Figure~\ref{fig:isingborn}(b) provides a specific choice of the parametrized quantum circuit proposed by Ref.~\cite{coyle2020born}, containing one fixed layer (Hadamard gates on all qubits) and two tunable layers (one with tunable two-qubit gates on all pairs of qubits and the other with tunable one-qubit gates on all qubits). Since tunable two-qubit gates mimic the Ising model, this model was named the ``Ising Born machine'' \cite{coyle2020born}. Although the original Ising Born machine was designed for generative modeling of classical data, it can be adapted to the general quantum learning task by removing the final quantum measurement layer in Fig.~\ref{fig:isingborn}(b). The main difference between the DIBoM and Ising Born machine lies in the number of tunable layers; the DIBoM can contain more than two tunable layers, thereby achieving higher expressive power than the Ising Born machine. In particular, the Ising Born machine is not capable of universal quantum computation, while the DIBoM is.

 \subsection{Hamiltonian learning}

Finally, the general quantum learning task is closely related to Hamiltonian learning \cite{cirstoiu2020variational,barison2021efficient}, a topic that has attracted a tremendous amount of interest recently. The correspondence is as follows: the quantum input $\ket{ \phi_{in} }$ corresponds to the initial quantum state of a system, and the quantum label $\ket{ \phi_{out} }$  corresponds to the quantum state of the system after evolving for a time $\delta t$. The hidden mapping $V$ can be associated with the time-evolution operator $e^{-i H \delta t}$ where $H$ is the Hamiltonian of the system. By approximating $V$ using QNNs, the Hamiltonian $H$ is learnt.

\section{Model}
\label{sec:model}

After reviewing related works, we now describe the DIBoM architecture. We first mathematically formulate the problem of quantum learning of quantum data in Sec.~\ref{sec:problemsetup}. Then in Sec.~\ref{sec:transformermodel}, we describe the DIBoM model which is targeted to this learning problem. Finally, we describe the training procedure of the DIBoM in Sec.~\ref{sec:quantumtraining}.

\subsection{Learning problem setup}
\label{sec:problemsetup}

We begin by introducing the quantum learning problem addressed by the DIBoM. Let $\mathcal{D}$ be an underlying distribution, and suppose we have $N$ pairs of training samples and labels $(\ket{\psi_i}, \ket{\phi_i}) \in \mathcal{D}$, where $1 \le i \le N$. For each pair, we are given $K$ copies of the input $\otimes_{i=1}^N(\ket{\psi_i}^{\otimes K}, \ket{\phi_i}^{\otimes K})$, as well as $M$ copies of test samples denoted by $\rho^{\otimes M}$. The goal is to generate model outputs for each test sample that closely approximate the corresponding test label. We use the infidelity to measure the similarity between two quantum states, and assume that the training and test data are independently sampled from $\mathcal{D}$. To illustrate, consider an example where $\ket{\psi_i}$ is a randomly generated $n$-qubit pure state and its corresponding label is $\ket{\phi_i}=V \ket{\psi_i}$. Here $V$ is a hidden $n$-qubit unitary that is independent of $i$ and unknown to the model. This example will also be used in the evaluation section. It should be noted that $\ket{\psi_i}$ and $\ket{\phi_i}$ may not be of the same dimension for general $\mathcal{D}$. 

\subsection{Model architecture}
\label{sec:transformermodel}
\begin{figure}[htb]
\centering \includegraphics[width=8cm]{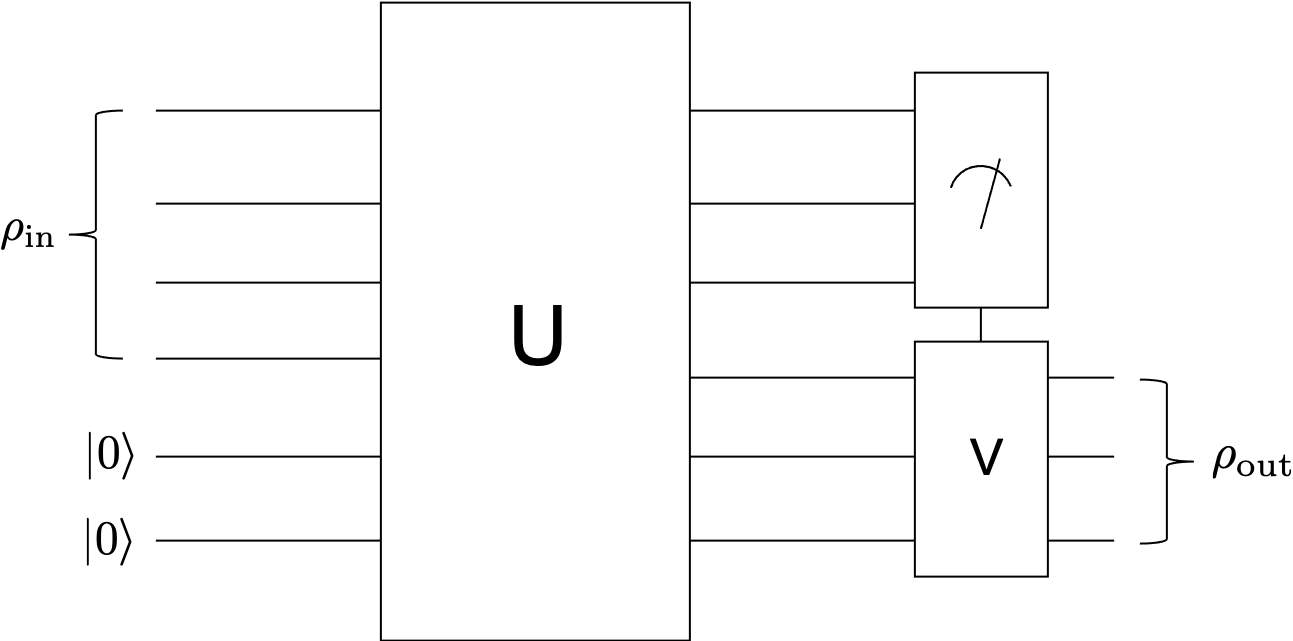}
\caption{The basic quantum structure. Here, $\rho_{in}$ and $\rho_{out}$ are its quantum input and quantum output respectively, $U$ is a unitary, and $V$ is a controlled unitary by the result of the measurement outcome.}
\label{fig:perceptron}
\end{figure}

After defining the learning problem, we now turn to the model of the DIBoM.  The DIBoM takes a quantum state as an input and outputs a quantum state as an output which may have different dimensions. It is based on a basic quantum structure which is illustrated in Fig.~\ref{fig:perceptron}. This basic quantum structure has three steps. First, the quantum input $\rho_{\mathrm{in}}$ together with a $k$-qubit ancilla $\ket{0}^{\otimes k}$ undergo a unitary transformation $U$ that produces an intermediate state 
\begin{equation}
 \rho_{\mathrm{inter1}} = U ( \rho_{\mathrm{in}} \otimes \ket{0}^{\otimes k} ) U^\dagger.
\end{equation}
Second, part of the joint quantum state is measured, resulting in outcome $j$.
Let $\rho_{\mathrm{inter2}}^j$ denote the post-measurement state conditioned on that the outcome is $j$.
Third, another unitary $V_j$, that can depend on the outcome $j$ in the second step, is applied to $\rho_{\mathrm{inter2}}^j$ to produce the output quantum state 
\begin{equation}
 \rho_{\mathrm{out}} = V_j \rho_{\mathrm{inter2}}^j V_j^\dagger.
\end{equation}

\begin{figure}[htb]
\centering \includegraphics[width=8.5cm]{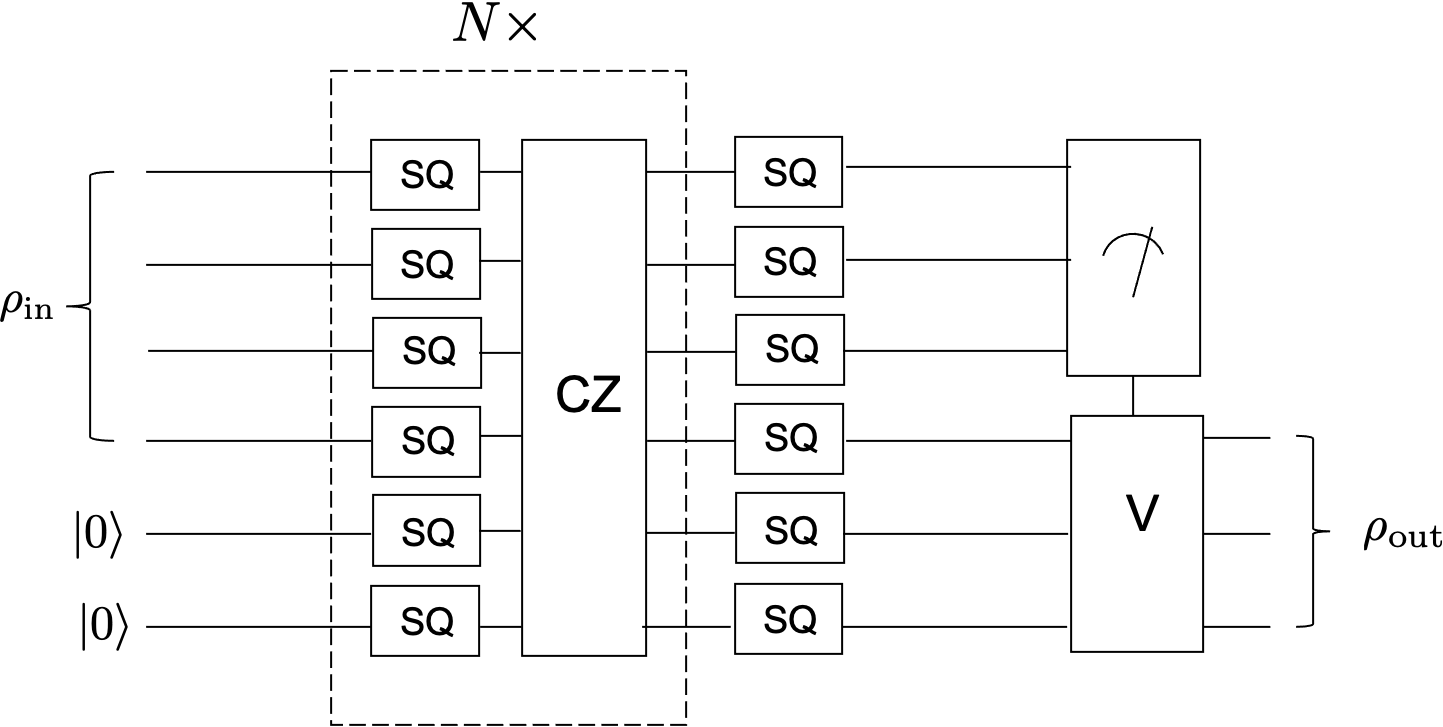}
\caption{The schematics of a deep Ising Born machine. Here, SQ denotes a tunable single-qubit gate;  CZ denotes generalized CZ gates between all pairs of qubits. The rest symbols have the same meanings as the basic quantum structure.}
\label{fig:main_model}
\end{figure}

With the basic quantum structure at hand, we are ready to define the DIBoM. Instead of applying the unitaries $U$ and $V_j$ which would consume exponential time to train, the DIBoM uses a stack of layers as a substitute for $U$ and $V_j$. The layers have two types. The first type consists of single-qubit rotations that are parametrized by $\alpha_j$ as follows:
\begin{equation}
 U_{\mathrm{SG}} = \exp \left[ i( \sum\limits_{j=1}^{3} \alpha_{j} \sigma_{j} ) \right],
\end{equation}
where $\sigma_1$, $\sigma_2$, $\sigma_3$ are Pauli matrices. 
The second type of layer applies a generalized CZ gate to all pairs of qubits:
\begin{equation}
 U_{\mathrm{CZ}} = \exp \left[ -i\pi( \sum\limits_{1\le j <k \le n} \beta_{jk} \ket{11}_{jk} \bra{11}_{jk} ) \right],
\end{equation}
where $\beta_{jk}$ is an arbitrary real number that interpolates smoothly between a CZ gate and an identity gate.
It is worth noting that if $\beta_{jk}=1$, a CZ gate is applied on qubits $j$ and $k$, and if $\beta_{jk}=0$, an identity gate is applied.
The DIBoM is constructed as $\prod_{j=L/2}^{1} ( U_{\mathrm{CZ}}^j U_{\mathrm{SG}} ^j )$ for $L$ even and $U_{\mathrm{SG}} ^{(L+1)/2} \prod_{j=(L-1)/2}^{1} ( U_{\mathrm{CZ}}^j U_{\mathrm{SG}} ^j )$ for $L$ odd, where $L$ is the total number of layers and $\prod_{j=1}^L U^j$ is short for $U^1\cdots U^L$. An illustration of the DIBoM is shown in Fig.~\ref{fig:main_model}.

\subsection{Training procedure}
\label{sec:quantumtraining}
 After presenting the DIBoM model, the next step is to discuss the training process. In this regard, two variants of the loss functions are considered. The first loss function, called the global loss function, has the form
\begin{equation}
\label{eq:lossfunction}
 \mathcal{L}_G =1 - \mathbb{E}_x \bra{\phi^x_\mathrm{out}} \rho^x_\mathrm{out}\ket{\phi^x_\mathrm{out}},
\end{equation}
where $\ket{\phi^x_\mathrm{out}}$ is the correct label, $ \rho^x_\mathrm{out}$ is the output of the DIBoM, and $ \mathbb{E}_x$ stands for the expectation over the random variable $x$. 
The intuition behind this loss function can be seen through some special cases.
If the correct label is identical to the model output, the loss is 0; otherwise, the loss is positive.
Hence, by minimizing the loss function, the model output converges to the correct label.
If the correct label is a mixed state $\sigma^x_\mathrm{out}$, the loss function can be easily generalized to
$ \mathcal{L}=1- \mathbb{E}_x F(\rho^x_\mathrm{out}, \sigma^x_\mathrm{out})$, where
 $F(\rho,\sigma):= \left [ \mathrm{tr} \sqrt{\rho^{1/2}\sigma\rho^{1/2}} \right]^2$.
 
The second loss function, called the local loss function, is given by
\begin{equation}
 \mathcal{L}_L = 1-\frac{1}{nN} \sum \limits_{x=1}^N  \sum \limits_{y=1}^n tr( ( \ket{\phi_x^{0}}_y\bra{\phi_x^{0}}_y \otimes I_{\bar{y}}  ) \rho_x^{0}(s) ) ,
 \label{eq:locallossfunction}
\end{equation}
where $\ket{\phi_x^{0}}_y$ is the $y$-th qubit of the input state $\ket{\phi_x^0}$, $I_{\bar{y}}$ denotes completely mixed states for all qubits except the $y$-th qubit,  $N$ is the number of samples, and $n$ is the number of qubits. Here, $\rho_x^{0}(s)$ represents the effective input which, when applied the unitary given by the current model, generates the correct quantum label. When applying the local loss function, it is assumed that the input quantum state $\ket{\phi_x^0}$ is a product state and has the form $\ket{\phi_x^0} = \ket{\phi_x^0}_1 \otimes \dots \otimes \ket{\phi_x^0}_n  $, where $n$ is the number of qubits. Note that no such assumption is made when applying the global loss function. 

\begin{figure}[htb]
\centering \includegraphics[width=6cm]{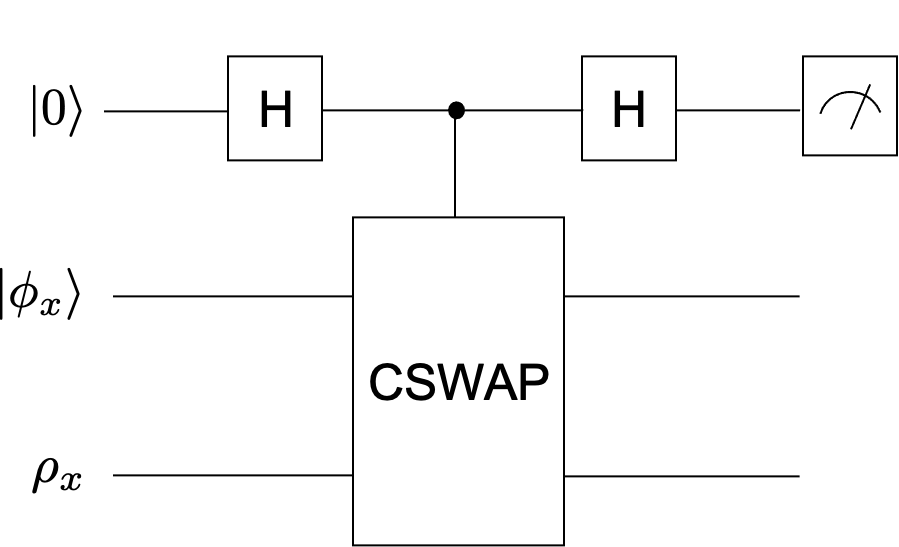}
\caption{Circuit to calculate the loss function. Here, $H$ is the Hadamard gate, $CSWAP$ is the controlled-SWAP gate, $\ket{\phi_x}$ is the ground-truth label, and $\rho_x$ is the model output.}
\label{fig:innerproduct}
\end{figure}

With the loss function defined (either $\mathcal{L}_G$ or $\mathcal{L}_L$, for simplicity denoted as  $\mathcal{L}$), we describe the procedure to train the network with quantum computers in two steps:
(i) calculate the loss function with quantum computers; (ii) update the parameters by performing gradient descent on the loss function.
For the first step, we first compute the quantity 
$\bra{\phi^x_\mathrm{out}} \rho^x_\mathrm{out}\ket{\phi^x_\mathrm{out}}$. To this end, we exploit the quantum circuit plotted in Fig.~\ref{fig:innerproduct} \cite{beer2020training}.
Through straightforward calculation, one can verify that this circuit takes $\ket{\phi^x_\mathrm{out}}$ and $ \rho^x_\mathrm{out}$ as inputs and outputs
$(1+\bra{\phi^x_\mathrm{out}} \rho^x_\mathrm{out}\ket{\phi^x_\mathrm{out}})/2$.
By a linear transformation and some classical computation, the loss function $\mathcal{L}$ is obtained. 

For the second step, we calculate the derivative of the loss function for all parameters  and update the parameters accordingly by performing gradient descent.
For a parameter $y^\mu$ (either $\alpha_j$ or $\beta_{jk}$) in any layer, we calculate its derivative by running the loss function calculation twice as follows,
\begin{equation}
\frac{ \delta \mathcal{L} }{ \delta y^\mu } = \frac{\mathcal{L}(y^\mu  )-\mathcal{L}(y^\mu - \epsilon )}{\epsilon},
\end{equation}
where $\epsilon$ is a small value. Note that this method assumes the availability of a high-precision quantum computer, as a noisy quantum computer may yield a derivative that is far from the true value. An alternative way would be calculating an analytic derivative directly with quantum computers, but this requires further investigation and is left as future work.

Then we update each parameter $y^\mu$ in the $k$-th iteration to minimize the loss function by the rule
\begin{equation}
y^\mu_{k+1} =y^\mu_{k} - \eta \frac{ \delta \mathcal{L} }{ \delta y^\mu_k },
\label{eq:naivegd}
\end{equation}
where $\eta$ is the learning rate. 
When $\eta$ is small enough, the loss function always decreases by the parameter update.
We note that the DIBoM is efficiently trainable as it has only a polynomial number of parameters.

The training strategy presented here does not aim to optimize the efficiency or computation cost of the training algorithm. Instead, we chose this strategy to evaluate the model's performance in terms of its converged loss. If the model can converge to a low test loss with some strategies, it is highly probable that the training strategy presented here will also result in a low test loss. 
This makes it quite ideal for testing the performance of the model in terms of its converged loss. Notably, there are training strategies such as the parameter-shift rule \cite{wierichs2022general} that can significantly reduce the number of quantum circuit evaluations, and gradient descent methods that offer faster convergence. For example, one may utilize Nesterov acceleration \cite{nesterov27method} which is also a first-order optimization method (utilizing only first-order derivatives) to speed up the convergence. The $k$-th iteration of the parameter $y^\mu$ in Nesterov acceleration has the form,
\begin{equation}
\begin{aligned}
x^\mu_{k+1} &=y^\mu_{k} +  \frac{k-1}{k+2} (y^\mu_{k}-y^\mu_{k-1}) , \\
y^\mu_{k+1} &=x^\mu_{k+1} - \eta \frac{ \delta \mathcal{L} }{ \delta x^\mu_{k+1} }.
\end{aligned}
\end{equation}
It can be shown that when $\eta = 1/\mathsf{L}$ where $\mathsf{L}$ is the Lipschitz constant of $\mathcal{L} $, the convergence rate by the above iteration rule is $O(1/k^2)$, quadratically better than $O(1/k)$ of Eq.~\eqref{eq:naivegd}. 

Second-order or higher-order optimization methods can offer further improvements in convergence by utilizing second-order derivatives $\partial^2 \mathcal{L} / \partial y^\mu \partial y^\nu$. However, the computational cost of each iteration step in second-order optimization methods is $O(N^2)$, in contrast to $O(N)$ of first-order optimization methods, where $N$ is the number of parameters. In classical neural networks, $N$ is usually on the order of $10^8$, making second-order optimization methods computationally too expensive. As a result, first-order optimization methods are generally preferred. Similarly, in QNNs, second-order optimization methods were considered inferior to first-order methods in cases where the learning problem required a large $N$ to solve. However, recent advances have shown that second-order methods can be substantially sped up \cite{gacon2021simultaneous}, making them a competitive alternative.

The training procedures presented above update all parameters simultaneously, and we refer to them as \emph{simultaneous training}. Another training method, known as \emph{layer-by-layer training} \cite{skolik2021layerwise}, offers an alternative approach. In each training step, the parameters of one layer are updated using gradient descent, while the parameters of all other layers are fixed. The layer to be trained can be selected in a round-robin manner, from layer $1$ to layer $L$ and then repeated, where $L$ is the number of layers. Alternatively, choosing the layer randomly is also a plausible approach. For the remainder of this paper, we will use the round-robin approach for layer-by-layer training, unless otherwise specified.

\section{Theoretical Analysis}
\label{sec:theory}

In this section, we conduct a theoretical analysis of the DIBoM architecture from three perspectives to gain insight into its properties. Specifically, we examine its flexibility with unequal input and output dimensions in Sec.~\ref{sec:inout}, its balance between expressive power and efficiency in Sec.~\ref{sec:power}, and compare it with other models in Sec.~\ref{sec:theocompare}.

\subsection{Input-output dimension}
\label{sec:inout}
We start by showing that the DIBoM can support unequal input and output dimensions.
This is due to its underlying structure, which was illustrated in Figure~\ref{fig:perceptron}. The DIBoM can accommodate a larger or smaller number of input qubits $m$ than output qubits $n$ by adjusting the number of ancilla qubits and the qubits to be measured. There are two cases to consider.  First, if $m<n$, an ancilla $\ket{0}^{\otimes (m-n)}$ can be used, and no measurement is required after the unitary $U$. Second, if $m>n$, no ancilla is needed, and a measurement can be performed on $m-n$ qubits after the unitary $U$.

As a result, quantum teleportation can be instantiated by the DIBoM as follows. We begin with a single-qubit quantum input $\rho_{in}$ and an ancilla in the state $\ket{00}$. A unitary operator $U$ is next applied to the system, which leaves the quantum input unchanged and entangles the ancilla into an EPR pair. A measurement is subsequently performed on both the quantum input and one of the qubits in the EPR pair. Based on the measurement outcome, an appropriate unitary operation is applied to the other qubit of the EPR pair. The result is the original quantum state being teleported to the quantum output $\rho_{out}$.

\subsection{Expressive power}
\label{sec:power}
Next we show another theoretical property of the DIBoM, namely its ability to perform universal quantum computation.
This is achieved by a reduction from a well-known result that  $2^{O(n)}$ layers of single-qubit gates and CZ gates suffice for universal quantum computation, where $n$ is the number of qubits \cite{nielsen2010quantum}. Note that a general circuit with $2^{O(n)}$ layers of single-qubit gates and CZ gates is inequivalent to the hardware-efficient QNN \cite{mcclean2018barren}. For example, Fig.~\ref{fig:illlus}(c), which belongs to the class of circuits with single-qubit gates and CZ gates, cannot be converted into the form of the hardware-efficient QNN \cite{mcclean2018barren}, while it can be turned into the form of DIBoM as we will see shortly. 

\begin{figure}[htb]
\centering \includegraphics[width=8.5cm]{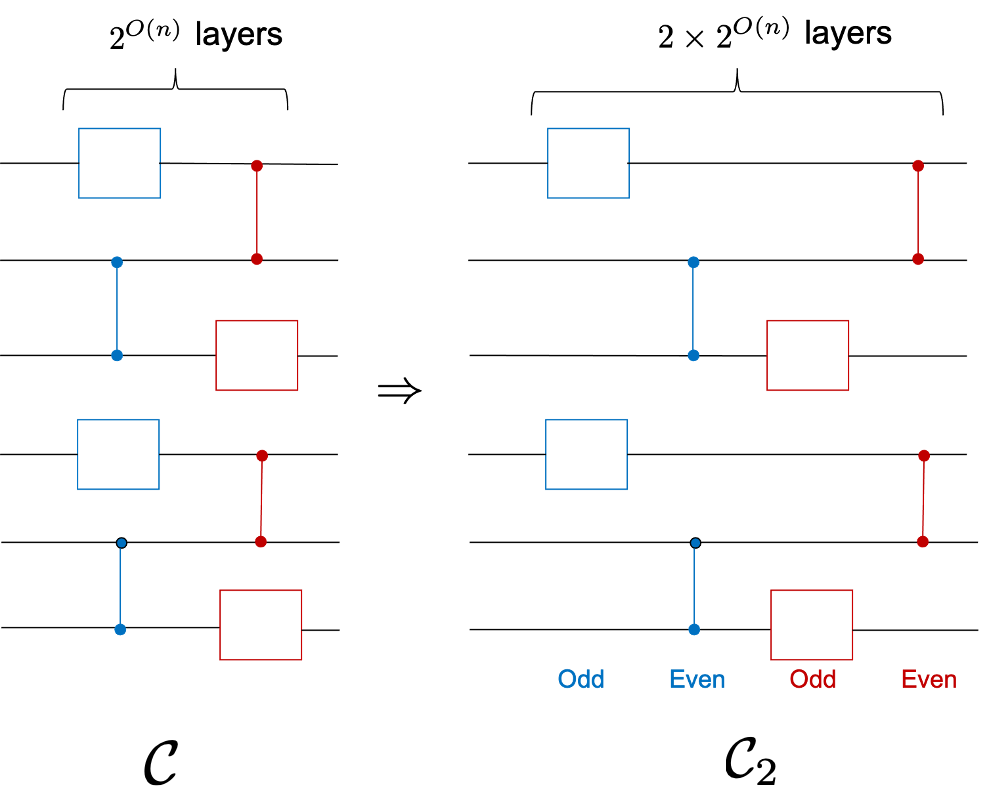}
\caption{A reduction from $\mathcal{C}$ to $\mathcal{C}_2$. Each layer of $\mathcal{C}$ consists of a mixture of one-qubit gates and CZ gates. In the reduction, every layer of $\mathcal{C}$ is split into two layers, the first involving only single-qubit gates and the second involving only CZ gates.  }
\label{fig:universalproof}
\end{figure}

Given a circuit $\mathcal{C}$ with $2^{O(n)}$ layers of single-qubit gates and CZ gates that approximates the desired unitary $U$ within an error of $\epsilon$, we convert it to the structure of a DIBoM in two steps.
\begin{enumerate}
\item In the first step, we split each layer of $\mathcal{C}$ into two layers, with the first layer containing only single-qubit gates and the second layer containing only CZ gates. The resulting circuit is called $\mathcal{C}_2$. 
\item In the second step, we fill missing single-qubit gates in the odd layer of $\mathcal{C}_2$ with identity single-qubit gates, and fill missing generalized CZ gates in the even layer of  $\mathcal{C}_2$ with identity two-qubit gates. 
 \end{enumerate}
 An illustration of this reduction is shown in Fig.~\ref{fig:universalproof}.
Hence a DIBoM with $2\times 2^{O(n)}=2^{O(n)}$ layers is capable of universal quantum computation. 

The representation power of the DIBoM forms a hierarchy that varies with the number of layers. At one end of the spectrum, when the DIBoM has a polynomial depth, it possesses a limited number of parameters, making it highly efficient. Conversely, at the other end of the spectrum, when the DIBoM has an exponential depth, it has the ability to approximate universal quantum computation with high precision. Hence the DIBoM balances the efficiency and the expressive power quite well.  

To quantitively compare the expressivity of the DIBoM and other QNN architectures, we propose an expressivity measure
\begin{equation}
E( \mathsf{A}) = \min_{U, \ket{\phi} } \max_\theta \left| \bra{\phi} \mathsf{A}(\theta)^\dagger U \ket{\phi} \right|,
\end{equation}
where $\mathsf{A}(\theta)$ and $\theta$ are the parametrized circuit and its parameters, $U$ is an arbitrary unitary, and $\ket{\phi}$ is an arbitrary pure quantum state.

To understand this measure, let us consider two special cases. First, when $\mathsf{A}(\theta)$ can recover any unitary, we can choose $\theta$ such that $\mathsf{A}(\theta) = U$ and hence $E( \mathsf{A}) =1$. Second, when $\mathsf{A}(\theta)$ is a fixed unitary, i.e. $\theta$ is empty, for an arbitrary $\ket{\phi}$, we can select a unitary $U$ such that $\mathsf{A}(\theta)\ket{\phi}$ and $U\ket{\phi}$ are orthogonal quantum states, hence $E( \mathsf{A}) =0$. Due to its close relationship with fidelity, we call this expressivity measure, \emph{fidelity-based expressivity} (FBE). Compared with other expressivity measures of QNNs, such as covering-number-based expressivity (CNBE) \cite{du2022efficient}, FBE has the advantages of having no extra parameters (CNBE has a parameter $\epsilon$), and having the range $[0,1]$ (CNBE is not upper bounded by a constant). In addition, FBE can be computed through continuous optimization, while some measures such as CNBE require discrete optimization which is usually harder to compute.

\begin{figure}[htb]
\centering \includegraphics[width=8.5cm]{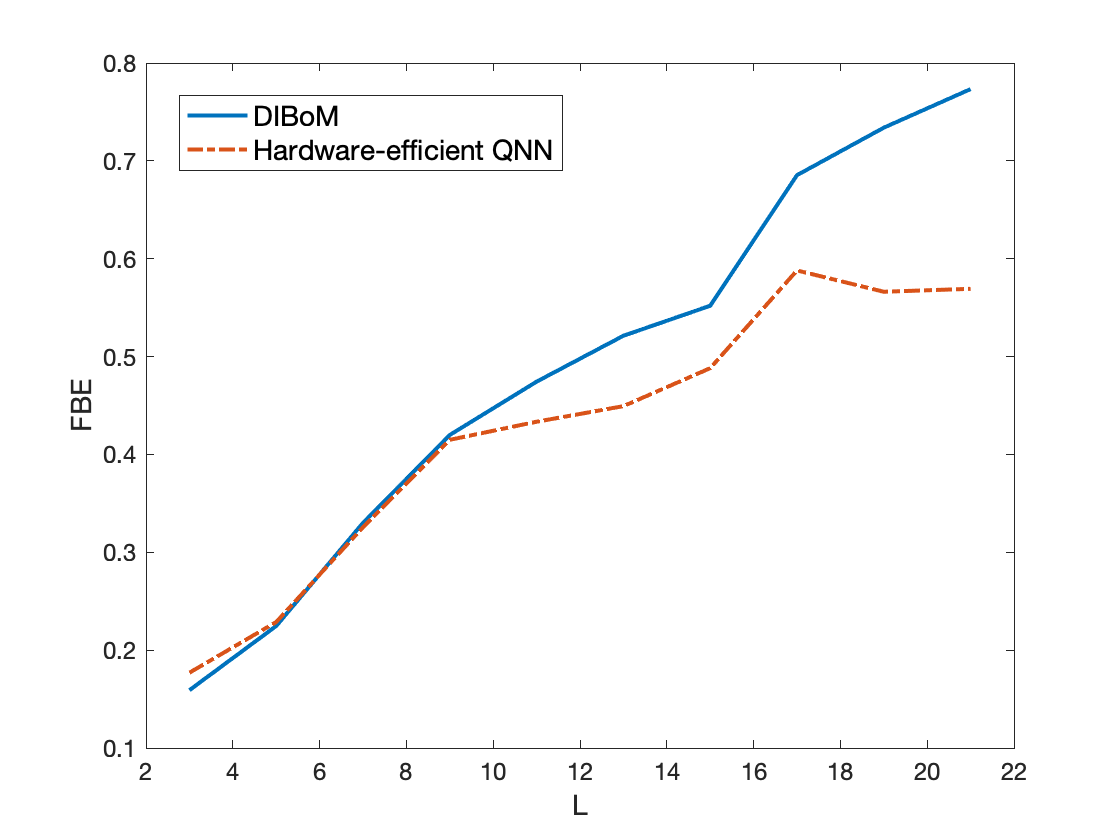}
\caption{ Fidelity-based expressivity (FBE) of the DIBoM and  hardware-efficient QNN as a function of the layer number $L$. }
\label{fig:expressivity}
\end{figure}

We now compare the expressive power of the DIBoM and hardware-efficient QNN through FBE. Specifically, we consider a three-qubit learning task and plot the FBE of the DIBoM and hardware-efficient QNN with $L$ layers as a function of $L$. (More details of the plot can be found in Appendix~\ref{appsec:fbe}.) Recall that the hardware-efficient QNN consists of alternating layers of single-qubit rotations and fixed CZ gates, while the DIBoM consists of alternating layers of single-qubit rotations and generalized CZ gates. Our results, as shown in Fig.~\ref{fig:expressivity}, indicate that the DIBoM outperforms the hardware-efficient QNN by a substantial margin when the layer number is the same. For instance, with 21 layers, the DIBoM achieves an FBE exceeding 0.77 (where higher values indicate superior performance), whereas the hardware-efficient QNN achieves only around 0.57. (Note however that DIBoM has more parameters than the hardware-efficient QNN with the same number of layers and hence this does not imply that DIBoM is strictly superior to the hardware-efficient QNN.)
Finally, the Ising Born machine, which corresponds to a DIBoM with $L=3$, exhibits significantly lower expressivity than the DIBoM with 21 layers, as shown in the figure.

\begin{figure}[htb]
\centering \includegraphics[width=8.5cm]{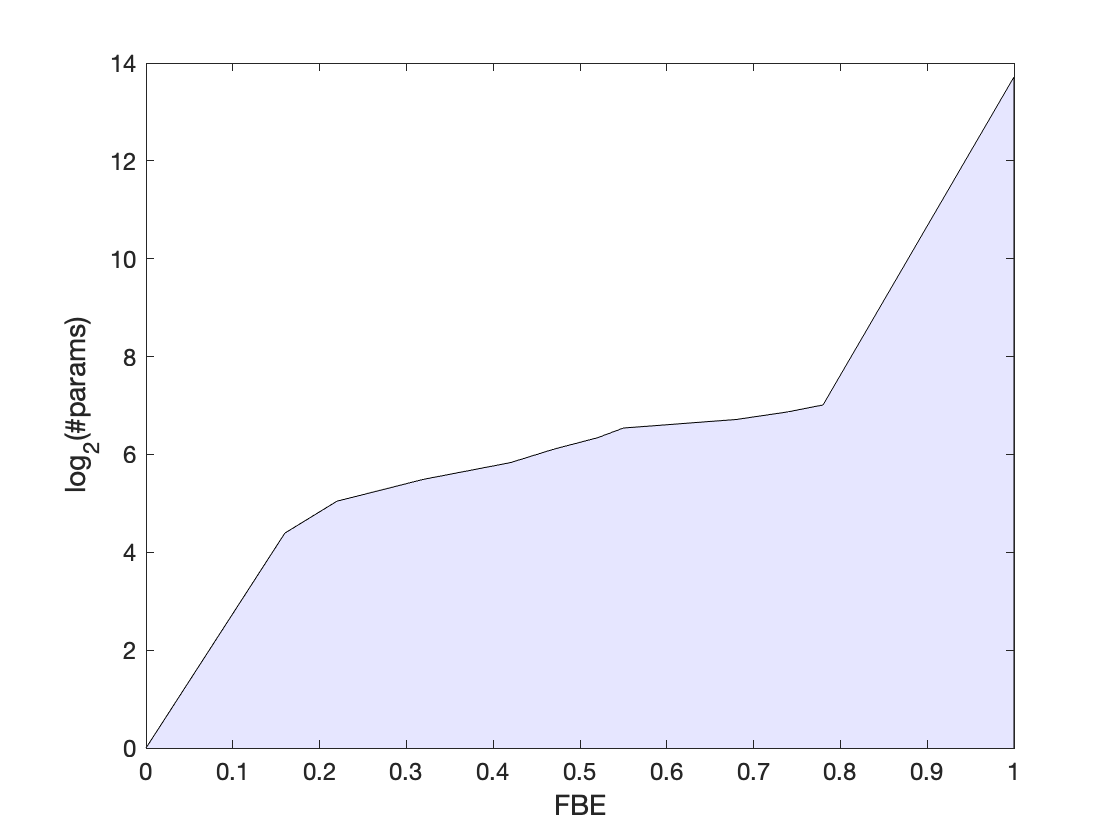}
\caption{ Quantitative characterization of the balance between efficiency and expressivity for a 3-qubit DIBoM. Here, the $y$ axis stands for efficiency which is measured as the logarithm of the number of parameters, and the $x$ axis stands for expressivity which is measured using fidelity-based expressivity (FBE).}
\label{fig:balance}
\end{figure}

The use of FBE also allows for a quantitative evaluation of the balance between efficiency and expressivity in DIBoMs. In Fig.~\ref{fig:balance}, we present such an evaluation for a 3-qubit DIBoM. Efficiency is quantified as the logarithm of the number of parameters, while expressivity is measured using FBE. The endpoints of the curve are obtained through theoretical analysis, where a 0-parameter DIBoM corresponds to an FBE value of 0 and a 13449-parameter DIBoM corresponds to an FBE value of 1. (The proof for the latter fact can be found in Appendix~\ref{appsec:3qubit}.) The remaining data points are computed numerically. The quality of the balance is characterized by the area of the purple region, with a smaller area indicating a better balance. It is worth noting that for some QNN architectures such as the hardware-efficient QNN, this area is not even guaranteed to be finite. In the figure, we show that through DIBoM, this area can be made finite. An interesting question for future work is how to achieve the smallest possible area of the purple region by optimizing over different QNN architectures.

\subsection{Comparisons with other models}
\label{sec:theocompare}

With the theoretical properties of the DIBoM at hand, we are ready to compare the DIBoM with other QNNs theoretically. First, we compare it to a basic QNN, as defined by Eq.~\eqref{eq:generalQNN}. The DIBoM has the advantage that its number of parameters is quadratic while a basic QNN has an exponential number of parameters. 

Second, we compare it to a dissipative QNN \cite{beer2020training}. In the case that the input and output have the same quantum dimension,  the DIBoM uses only half the number of qubits required by a dissipative QNN. In addition, the number of parameters that a DIBoM uses is exponentially smaller than that of a dissipative QNN.

\section{Empirical evaluation setup}
\label{sec:simulationsetup}

To further investigate the properties of the DIBoM architecture, we conduct an extensive empirical evaluation of the DIBoM in this and the following sections. In this section, we present the setup of the evaluation, while the results of the evaluation are presented in the next section.
The setup consists of two parts. In the first part, we describe the synthetic dataset that is used in the evaluation of the DIBoM. In the second part, we give the classical simulation for the training of the DIBoM.   

\subsection{Dataset}
\label{sec:data}

We start with the construction of the synthetic datasets used in the empirical evaluation.
The samples in each dataset are of the form $\ket{\phi_x^{in}}$ where $x=1,\dots,N$ and 
$N$ is the number of samples.
Each sample is associated with a corresponding label $\ket{\phi_x^{out}}=V\ket{\phi_x^{in}}$ where $x=1,\dots,N$. The unitary $V$ is referred to as the \emph{intrinsic unitary} of the  data and is hidden from the training models.
If not otherwise specified, the samples are 2-qubit states and chosen randomly. 
In our main experiment, we generate a total of 20 samples, which are randomly divided into equal-sized training and test sets (50:50).

\subsection{Classical simulation of training}
\label{sec:simultaneous}

With the dataset in place, we next describe how to evaluate the performance of a DIBoM on the dataset.
Due to the lack of a quantum computer, we classically simulate the 
training procedure of the DIBoM and examine the result. 
In the following, we describe the classical simulation of the network training for a DIBoM.
Let $L+1$ be the number of layers in the network,  where layer 0 is the input layer and layer $L$ the output layer. The transition from layer $l-1$ to layer $l$ is given by
\begin{eqnarray}
\rho_x^l(s) &=&  U^l (s) \rho_x^{l-1}(s)   {U^l}^\dagger(s) ,  
\end{eqnarray}
where $s$ is any parameter of the model (such as $\alpha_j$ or $\beta_{jk}$) and $U^l (s)$ is the unitary in layer $l$.
The loss function is computed as $\mathcal{L}(s)=1-C(s)$, where
\begin{equation}
C(s) = \frac{1}{N} \sum \limits_{x=1}^N \bra{\phi_x^{L}} \rho_x^{L}(s) \ket{\phi_x^{L}},
\end{equation}
with $N$ being the number of data points, $\rho_x^{L}(s)$ denoting the output state of the network, and $\ket{\phi_x^L}$ denoting the label.

In each iteration, the unitaries in the network are updated by 
$U^l (s+\epsilon) = e^{i\epsilon K^l(s)} U^l(s)$.
Therefore, the network training is equivalent to obtaining $K^l(s)$ in each iteration. 
To this end, we first calculate the derivative of $C$ with respect to the parameter $s$, which is
\begin{equation}
\frac{dC}{ds} = \lim\limits_{\epsilon \to 0} \frac{ C(s+\epsilon)-C(s)}{ \epsilon},
\end{equation}
where $\epsilon$ is a small positive number.
To evaluate this derivative, we first obtain the expression of $C(s+\epsilon)$. For the parameter $s+\epsilon$, the input quantum state stays unchanged as
$\rho_x^{0} (s+\epsilon)= \rho_x^{0} = \ket{\phi_x^{0}}\bra{\phi_x^{0}}$.
The quantum output, by the composition of layers, is however changed and can be expressed as
\begin{eqnarray*}
 \rho^{L}_x (s+\epsilon) &= &\prod_{l=L}^1 e^{i\epsilon K^{l}(s)} U^{l}(s) \rho_x^{0}   \prod_{l=1}^{L} {U^{l}}^\dagger (s) e^{-i\epsilon K^{l}(s)} . \nonumber
\end{eqnarray*}
We then substitute the updated quantum output corresponding to the parameter $s+\epsilon$ into the derivative of $C$, obtaining
\begin{equation}
\frac{dC}{ds} = \frac{i}{N} \sum\limits_x \mathrm{tr} ( \sum_{l=L}^1  M^{l}(s) K^{l}(s) ), 
 \label{eq:derivative}
\end{equation}
where $N$ is the number of samples, $\mathrm{tr}$ denotes the trace operation, and $M^{l}(s)$ is defined as
\begin{eqnarray*}
M^{l}(s)& =&  [  \prod_{j=l}^1 U^{j}(s)  \rho_x^0   \prod_{j=1}^l {U^{j}}^\dagger (s) ,   \\
 && \quad \prod_{j=l+1}^L{U^{j}}^\dagger (s)   \ket{\phi_x^L}\bra{\phi_x^L} \prod_{j=L}^{l+1} U^{j} (s) ].
\end{eqnarray*}
Here, $[\cdot,\cdot]$ denotes the commutator operation. The derivation of Eq.~\eqref{eq:derivative} is given in Appendix \ref{appsec:derivation}.

To maximize the increase of $C$, $K^l(s)$ should be chosen such that $dC/ds$ is maximized. For this purpose, we consider $K^l(s)$ which corresponds to a general $n$-qubit unitary $U^l(s)$.
To avoid overfitting, we impose regularization on the parameters. 
Specifically, we regularize the parameters $K_{\alpha_1, \cdots, \alpha_n}^l(s)$ which are defined as
\begin{eqnarray}
K^l(s) &=& \sum \limits_{\alpha_1,  \cdots, \alpha_{n}} K_{\alpha_1, \cdots, \alpha_{n}}^l(s)  (\otimes_{k=1}^n \sigma^{\alpha_k }).
\end{eqnarray}
Hence, the combined objective (which both maximizes the derivative of $C$ and minimizes the change of network parameters) to be maximized is 
\begin{eqnarray}
C_2 &=& \frac{dC}{ds} - \lambda \sum\limits_{\alpha_i}  K_{\alpha_1, \cdots, \alpha_n}^l(s) ^2  \nonumber \\
&=& \frac{i}{N} \sum\limits_x \mathrm{tr}(  \sum_{l=L}^1 M^{l}(s) K^{l}(s) )  - \lambda \sum\limits_{\alpha_j}  K_{\alpha_1, \cdots, \alpha_n}^l(s) ^2 \nonumber \\
&=&  \frac{i}{N} \sum\limits_x \mathrm{tr}_{\alpha_1, \dots, \alpha_n} (\mathrm{tr}_{rest}(  \sum_{l=L}^1  M^{l}(s) K^{l}(s) )) \nonumber \\
&&  - \lambda \sum\limits_{\alpha_j}  K_{\alpha_1, \cdots, \alpha_n}^l(s) ^2.  
\end{eqnarray}

To maximize $C_2$, we calculate its derivative with respect to $ K_{\alpha_1, \cdots, \alpha_n}^l(s)$ as 
$i  \sum_x \mathrm{tr}_{\alpha_1, \cdots, \alpha_n} (\mathrm{tr}_{rest}(  M^{l}(s) ) (\otimes_{k=1}^n \sigma^{\alpha_k }) )/N - 2\lambda  K_{\alpha_1, \cdots, \alpha_n}^l(s)$.
By setting it to 0 and solving for $K_{\alpha_1,\cdots,\alpha_n}^l(s)$, we obtain
\begin{equation*}
K_{\alpha_1, \cdots, \alpha_n}^l(s)= \frac{i}{2N\lambda} \sum\limits_x \mathrm{tr}_{\alpha_1, \cdots, \alpha_n} (\mathrm{tr}_{rest}(  M^{l}(s) )(\otimes_{k=1}^n \sigma^{\alpha_k }).
\end{equation*}
There is a caveat that $C_2$ might be always negative, in which case we should not update $s$. To ensure that the solution of $dC_2/dK_{\alpha_1,\dots,\alpha_n}^l=0$ results in an increase in $C$, we explicitly check the value of $C$ before updating $s$.
We substitute the obtained value of $K_{\alpha_1, \dots, \alpha_n}^l(s)$ back to $K^l(s)$ and obtain
\begin{eqnarray}
K^l(s) 
&=& \frac{i}{2N\lambda} \sum \limits_{\alpha_1, \cdots, \alpha_n}  \sum\limits_x  \mathrm{tr}_{\alpha_1, \cdots, \alpha_n} (\mathrm{tr}_{rest}(  M^{l}(s) )  \nonumber \\
&& (\otimes_{k=1}^n \sigma^{\alpha_k }) ) (\otimes_{k=1}^n \sigma^{\alpha_k }) \nonumber \\
&=& i 2^{n} \sum\limits_x \mathrm{tr}_{rest}(  M^{l}(s) ) / (N\lambda). 
\end{eqnarray}
Finally the unitary $U^l$ is updated by 
\begin{equation*}
 U^l (s+\epsilon) = \exp(-\epsilon 2^{n} \sum\limits_x \mathrm{tr}_{rest}(  M^{l}(s) ) / (N\lambda)) U^l(s).
\end{equation*}

Now we consider the specific unitaries used by the DIBoM which can be categorized into three cases:
\begin{enumerate}
\item The first case is $U_{SG}^j$, which is a single-qubit unitary on the qubit $j$. The corresponding $K$ for this unitary is
$K^l(s)  = \sum_{\alpha=0}^3 K^l_\alpha(s) \sigma^\alpha$.
To obtain the optimal $K^l_\alpha(s)$, we set $d C_2/d K^l_\alpha(s)=0$ and obtain
\begin{equation}
K^l_\alpha(s)=  \frac{i}{2N\lambda} \sum\limits_x  \mathrm{tr}_{j} (\mathrm{tr}_{rest}(   M^{l}(s) )  \sigma^{\alpha }  ),
\end{equation}
where ``rest'' denotes qubits other than qubit $j$.
Substituting the expression back into $K^l(s) $, we have
\begin{eqnarray}
K^l(s) 
&=&   \frac{i}{2N\lambda} \sum \limits_{\alpha}   \sum\limits_x   \mathrm{tr}_{j} (\mathrm{tr}_{rest}(   M^{l}(s) )  \sigma^{\alpha }  )   \sigma^{\alpha } \nonumber  \\
&=&  i \sum\limits_x \mathrm{tr}_{rest}(M^{l}(s) )/(N\lambda).
\end{eqnarray}
Therefore the unitary is updated as $U^l (s+\epsilon) = \exp( i \epsilon  K^l(s)  ) U^l(s)$.

\item  The second case is a product of single-qubit gates $U_{SG}^\otimes$, the corresponding $K$ of which has the form
\begin{equation}
K^l(s)  = \sum_{j=1}^n \sum_{\alpha=0}^3 K^l_{j,\alpha}(s) \sigma^\alpha_j.
\end{equation}
By letting $d C_2/d K^l_{j,\alpha}(s)=0$, we obtain
\begin{equation}
K^l_{j,\alpha}(s)=  \frac{i}{2N\lambda} \sum\limits_x  \mathrm{tr}_{j} (\mathrm{tr}_{[n]\backslash\{j\}}(   M^{l}(s) )  \sigma^{\alpha }_j  ),
\end{equation}
where ${[n]\backslash\{j\}}$ refers to all qubits except qubit $j$.
Substituting this expression back into $K^l(s) $, we have
\begin{eqnarray}
K^l(s) 
&=&   \frac{i}{2N\lambda} \sum\limits_{j=1}^n\sum \limits_{\alpha}   \sum\limits_x   \mathrm{tr}_{j} (\mathrm{tr}_{[n]\backslash\{j\}}(   M^{l}(s) )  \sigma^{\alpha }_j  )   \sigma^{\alpha }_j \nonumber   \\
&=& i \sum\limits_{j=1}^n\sum\limits_x \mathrm{tr}_{[n]\backslash\{j\}}(   M^{l}(s) )/ (N\lambda).
\end{eqnarray}
Therefore the unitary is updated as
 $U^l (s+\epsilon) = \exp( i \epsilon K^l(s)  ) U^l(s)$.

\item  The third case is the collection of generalized CZ gates on all pairs of qubits $U_{CZ}$. Its corresponding  $K$ is
$K^l(s)  = \sum_{1\le j<k\le n} K^l_{jk}(s) \ket{11}_{jk}\bra{11}$.
By setting $d C_2/d K^l_{jk}(s) =0$, we obtain
\begin{equation}
K^l_{jk}(s)=  \frac{i}{2N\lambda} \sum\limits_x  \mathrm{tr}_{j,k}(\mathrm{tr}_{[n]\backslash\{j,k\}}   M^{l}(s) )  \ket{11}_{jk}\bra{11} ),
\end{equation}
where ${[n]\backslash\{j,k\}} $ refers to the set of qubits excluding qubits $j$ and $k$.
Substituting this expression back into $K^l(s)$, we have
\begin{eqnarray*}
K^l(s) &=&   \frac{i}{2N\lambda} \sum \limits_{j,k}   \sum\limits_x   \mathrm{tr}_{j,k}(\mathrm{tr}_{[n]\backslash\{j,k\}} (   M^{l}(s) )  \ket{11}_{jk}  \nonumber \\
&&\bra{11} )     \ket{11}_{jk}\bra{11}  \nonumber \\
&=&  i [\sum\limits_x \sum \limits_{j,k}   \mathrm{tr}_{j,k}(\mathrm{tr}_{[n]\backslash\{j,k\}} (   M^{l}(s) )  \ket{11}_{jk}\bra{11} )   \nonumber \\
  &&   \ket{11}_{jk}\bra{11}] / (2 N\lambda).
\end{eqnarray*}
Therefore the unitary is updated as $U^l (s+\epsilon) = \exp( i \epsilon K^l(s)  ) U^l(s)$.
\end{enumerate}

Three final remarks are in order. First, the hyperparameter $\lambda$ is set to $0.5$ in the simulation unless otherwise stated. Second, the classical simulation of the gradient descent for each parameter $s$ in layer-by-layer training is identical to the simultaneous training method. Third, the classical simulation for training controlled unitaries $V_j$ is similar, and the specifics are deferred to Appendix~\ref{app:simulate_control}.

\section{Empirical evaluation results}
\label{sec:simulationresult}

Having presenting the simulation setup, this section proceeds to present the empirical evaluation results. We first empirically compare the performance of DIBoM with other QNNs in Sec.~\ref{sec:trainperform}. Next, we conduct an ablation study on the DIBoM in Sec.~\ref{sec:ablation} to investigate the individual components of the model. Then in Sec.~\ref{sec:sensitivity}, we assess the sensitivity of the performance of the DIBoM to its various parameters. Additionally, in Sec.~\ref{sec:robustness}, we analyze the robustness of the DIBoM to noise. Finally, in Sec.~\ref{sec:barren}, we mitigate the barren plateau issue in the training of the DIBoM. Some auxiliary details pertaining to the construction of the datasets used in this section are presented in Appendix~\ref{appsec:detail}. 

\subsection{Performance comparison}
\label{sec:trainperform}

To begin, we analyze the training performance of the DIBoM and compare it to that of other models, considering both the converged loss and the model complexity. Additionally, we explore two different training methods and evaluate the gap between training and test performance. Moreover, we plot the optimization landscape of the DIBoM and dissipative QNN to gain a deeper understanding of their respective training processes.

\begin{figure}[htb]
\centering \includegraphics[width=8.5cm]{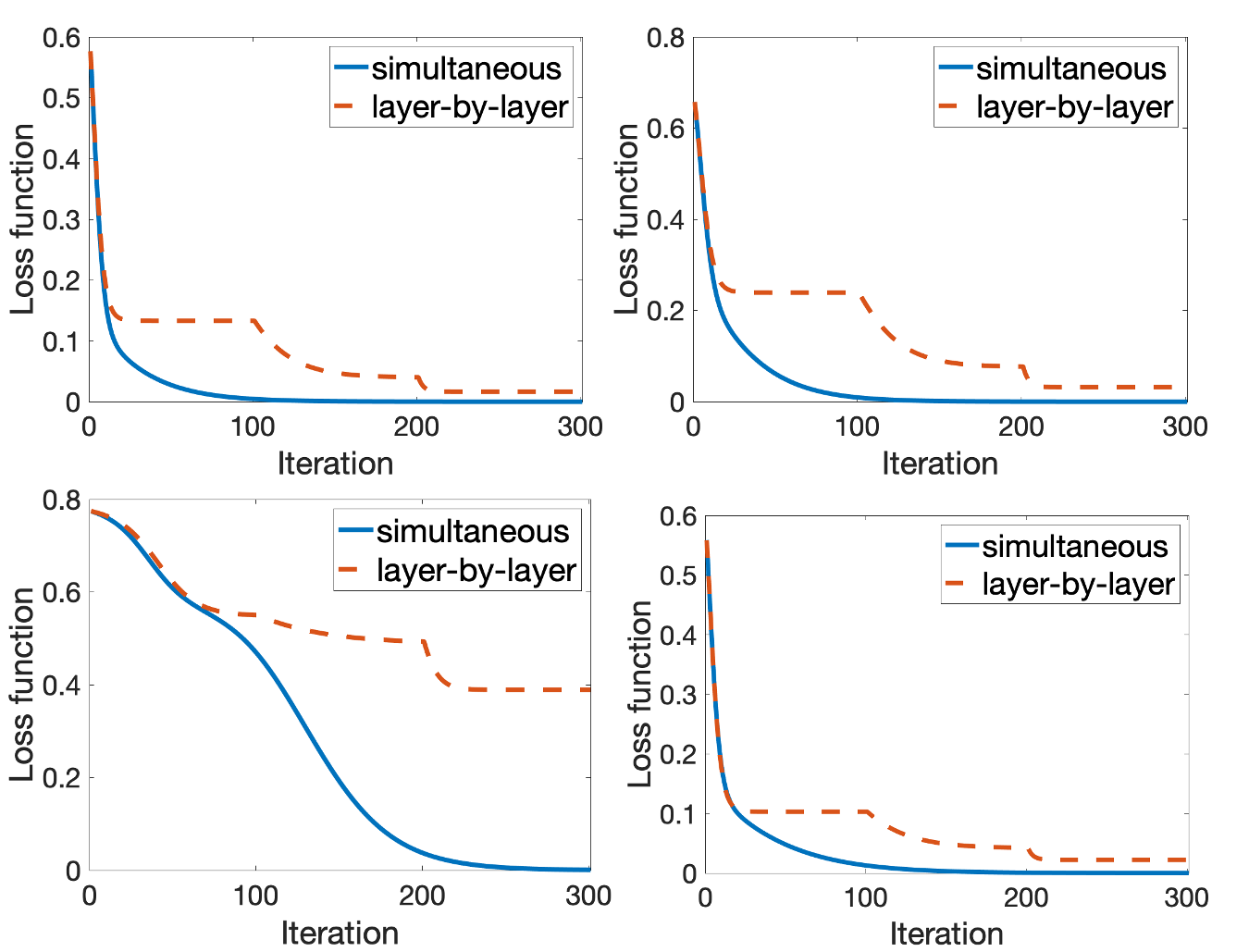}
\caption{Four learning curves for a DIBoM with two training methods. 
}
\label{fig:plotHybrid}
\end{figure}

We first test the simultaneous training and layer-by-layer training methods and display four training results with different datasets in Fig.~\ref{fig:plotHybrid}. The results indicate that the model's loss converges to 0. Comparing the two training methods, we observe that layer-by-layer training consistently performs worse than simultaneous training. Thus, we will solely utilize simultaneous training in future simulations.

\begin{figure}[htb]
\centering \includegraphics[width=8.5 cm]{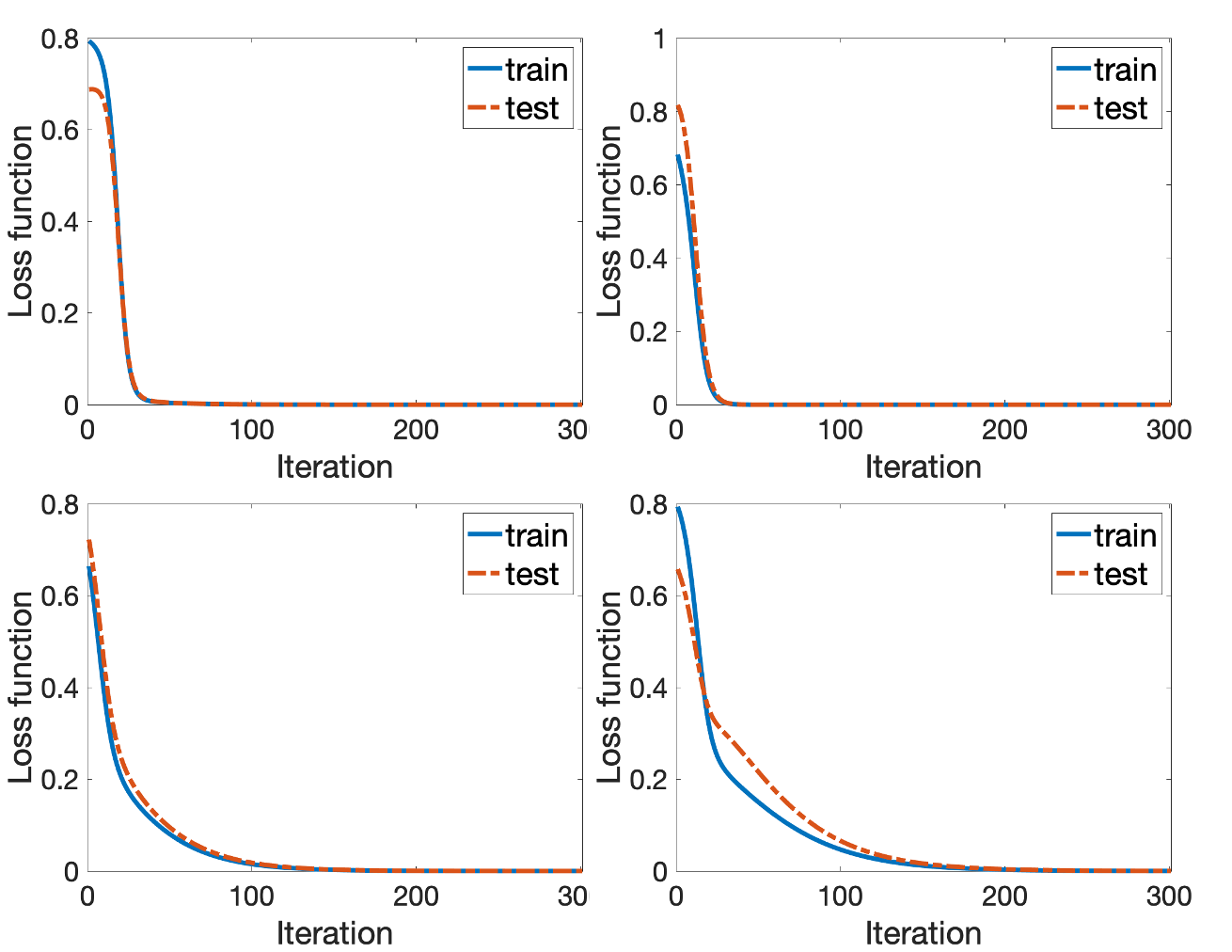}
\caption{Four training and test learning curves for a DIBoM. 
}
\label{fig:compare}
\end{figure}

We next examine the DIBoM's prediction accuracy on the test set and assess its gap with the training accuracy, as depicted in Fig.~\ref{fig:compare}. Notably, the training loss and test loss are nearly identical, with the test loss occasionally being smaller than the training loss. This suggests that statistical fluctuations rather than generalization errors may cause the deviation between the training and test losses. Given the close proximity of the two losses, we will exclusively evaluate the test loss in subsequent simulations.

\begin{figure}[htb]
\centering \includegraphics[width=8.5 cm]{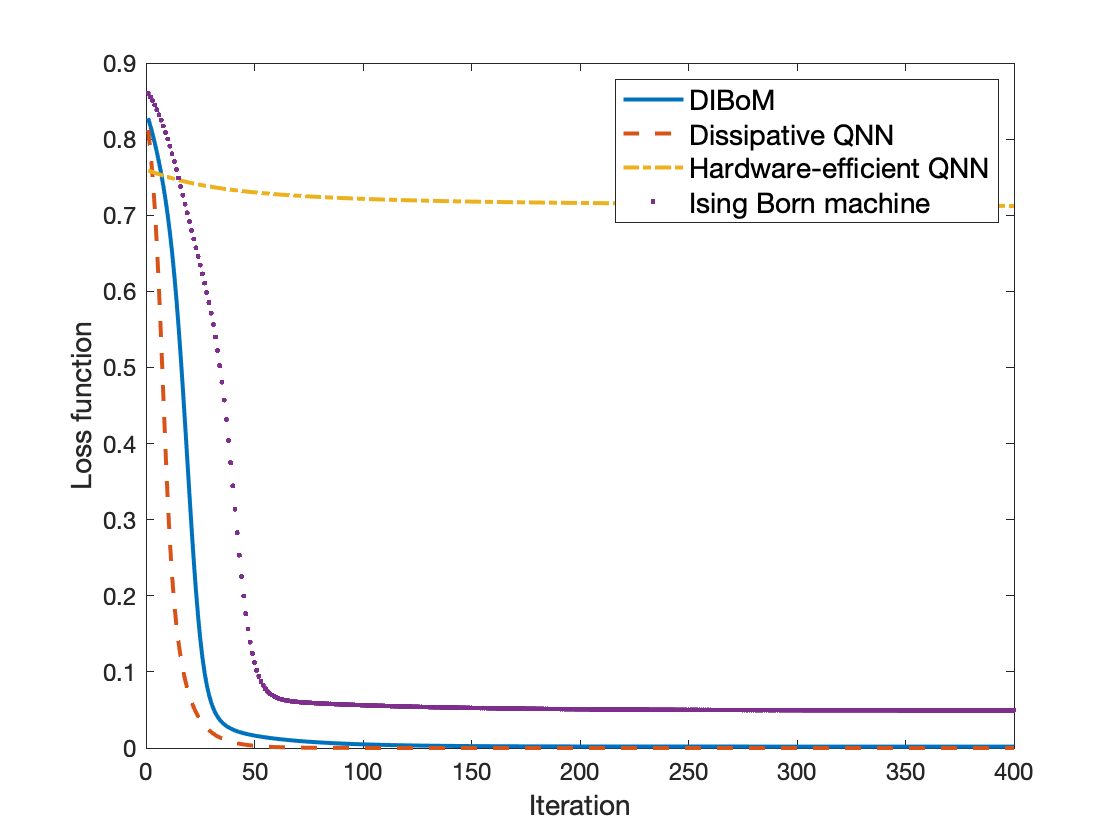}
\caption{Comparison of a DIBoM with a dissipative QNN,  a hardware-efficient QNN and an Ising Born machine in terms of the performance for a structured unitary $V$. 
}
\label{fig:baseline}
\end{figure}

\begin{figure}[htb]
\centering \includegraphics[width=8.5 cm]{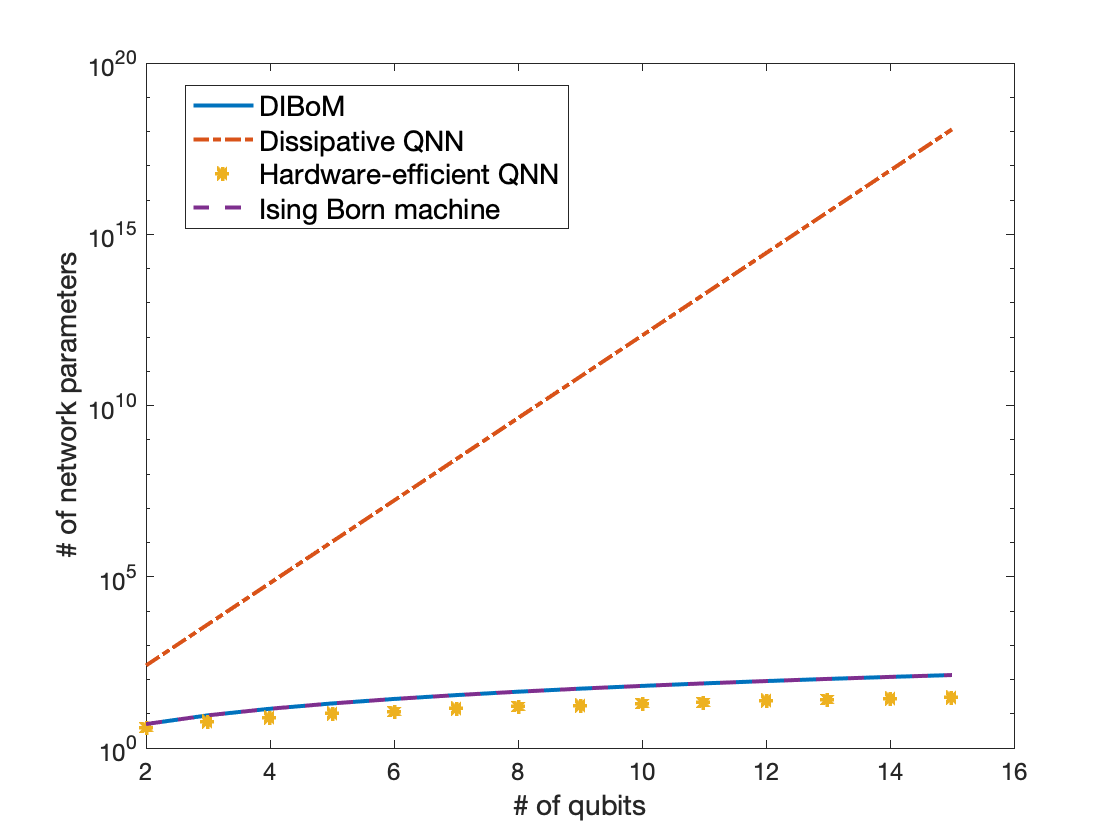}
\caption{Comparison of a DIBoM  with a dissipative QNN,  a hardware-efficient QNN and an Ising Born machine in terms of the number of model parameters. 
}
\label{fig:paramCompare}
\end{figure}

We then compare the DIBoM with three other QNNs: a hardware-efficient QNN \cite{mcclean2018barren}, a dissipative QNN \cite{beer2020training} and an Ising Born machine \cite{coyle2020born}.  In the simulation, we set the number of qubits to be 2, the DIBoM to contain five alternating layers of tunable single-qubit gates and tunable two-qubit gates, the hardware-efficient QNN to have the same structure as the DIBoM but with all tunable two-qubit gates replaced by CZ gates, and the Ising Born machine to contain a layer of tunable two-qubit gates followed by a layer of tunable single-qubit gates.
The intrinsic unitary $V$ of this simulation is restricted to have the same structure as the DIBoM but with unknown parameters. We compare the models based on two criteria. First, we compare them in terms of their performance, as shown in Fig.~\ref{fig:baseline}. We observe that both the dissipative QNN and the DIBoM reach zero loss while the hardware-efficient QNN and the Ising Born machine do not. This could be attributed to the limited expressive power of the hardware-efficient QNN and the Ising Born machine. We also observe that the  dissipative QNN converges faster than the DIBoM. Second, we compare them in terms of the number of
model parameters in Fig.~\ref{fig:paramCompare}. We observe that  the
dissipative QNN requires significantly more parameters than the other three models. As the number of qubits increases, the ratio of the number of parameters of the DIBoM to that of the dissipative QNN tends to 0. Hence, there is a tradeoff between performance and the number of parameters.

\begin{figure}[htb]
\centering \includegraphics[width=8.5 cm]{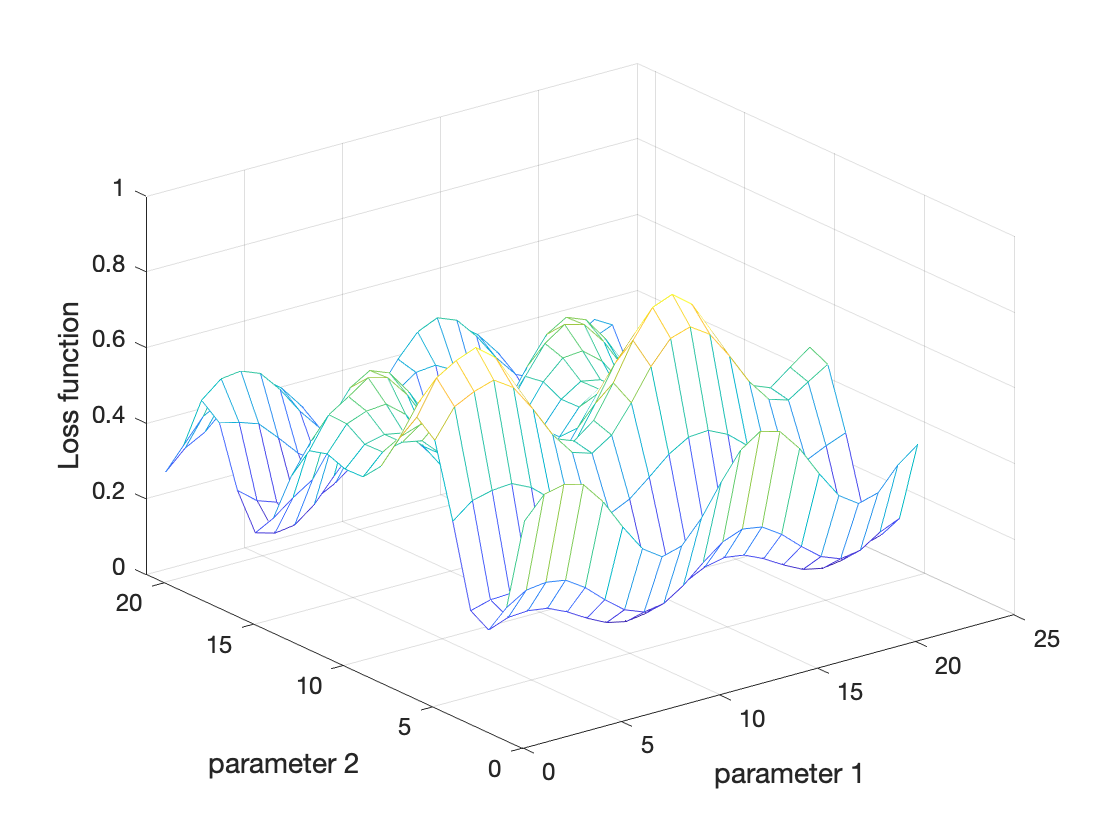}
\caption{Optimization landscape of a DIBoM with two parameters varying and other parameters fixed. 
 }
\label{fig:landscape}
\end{figure}

\begin{figure}[htb]
\centering \includegraphics[width=8.5 cm]{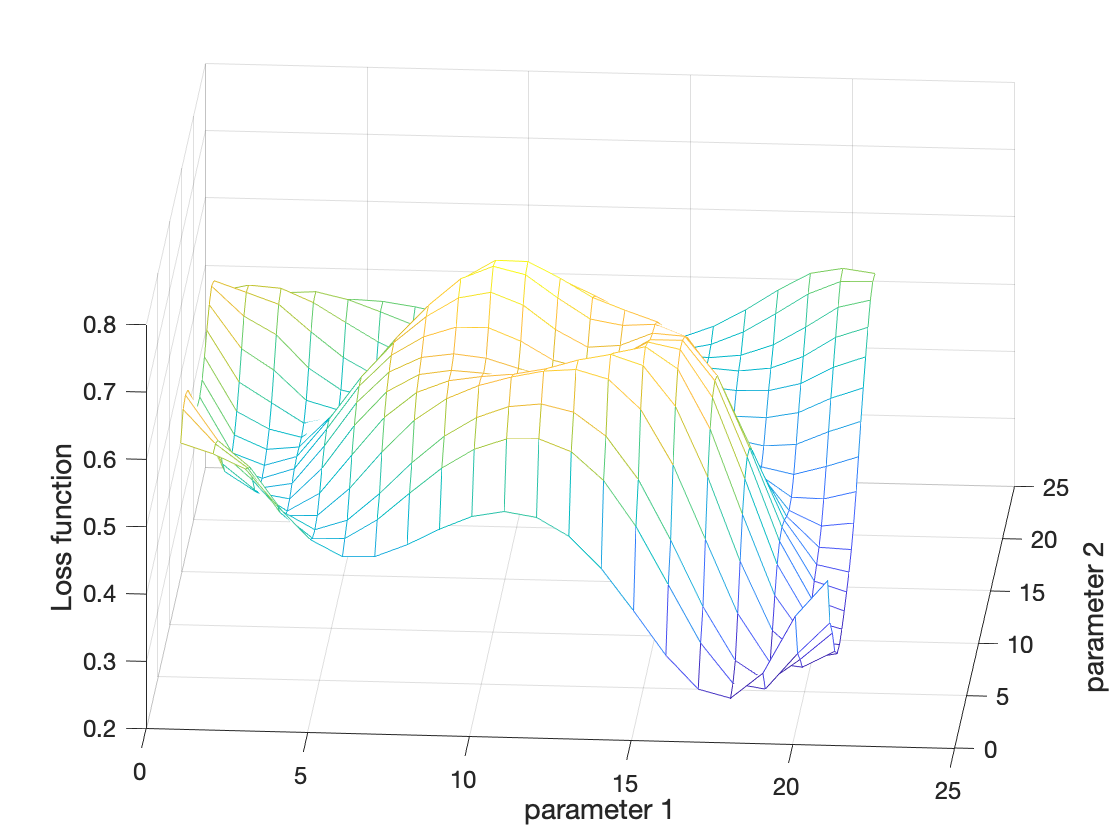}
\caption{Optimization landscape of a dissipative QNN with two parameters varying and other parameters fixed. 
}
\label{fig:dissipative_landscape}
\end{figure}

To investigate why the DIBoM and dissipative QNN can achieve zero loss, we plot the optimization 
landscapes of them with two parameters varying and other parameters fixed. 
The result of the DIBoM is shown in Fig.~\ref{fig:landscape}, where one parameter to be changed is from the single-qubit unitary gate (parameter 1) and the other parameter to be changed is from the generalized CZ gate (parameter 2). Despite the highly non-convex landscape, all local minima are global minima, explaining why the DIBoM can always converge to zero loss.  
The result of the dissipative QNN is shown 
in Fig.~\ref{fig:dissipative_landscape}. It can be seen that there is no flat region in the landscape and 
this explains the fast convergence of a dissipative network. However, there exists a local minimum in the middle of the figure which 
does not coincide with the global minimum. Hence, whether dissipative networks can always be trained to reach a global minimum requires further investigation.

\begin{figure}[htb]
\centering \includegraphics[width=8.5 cm]{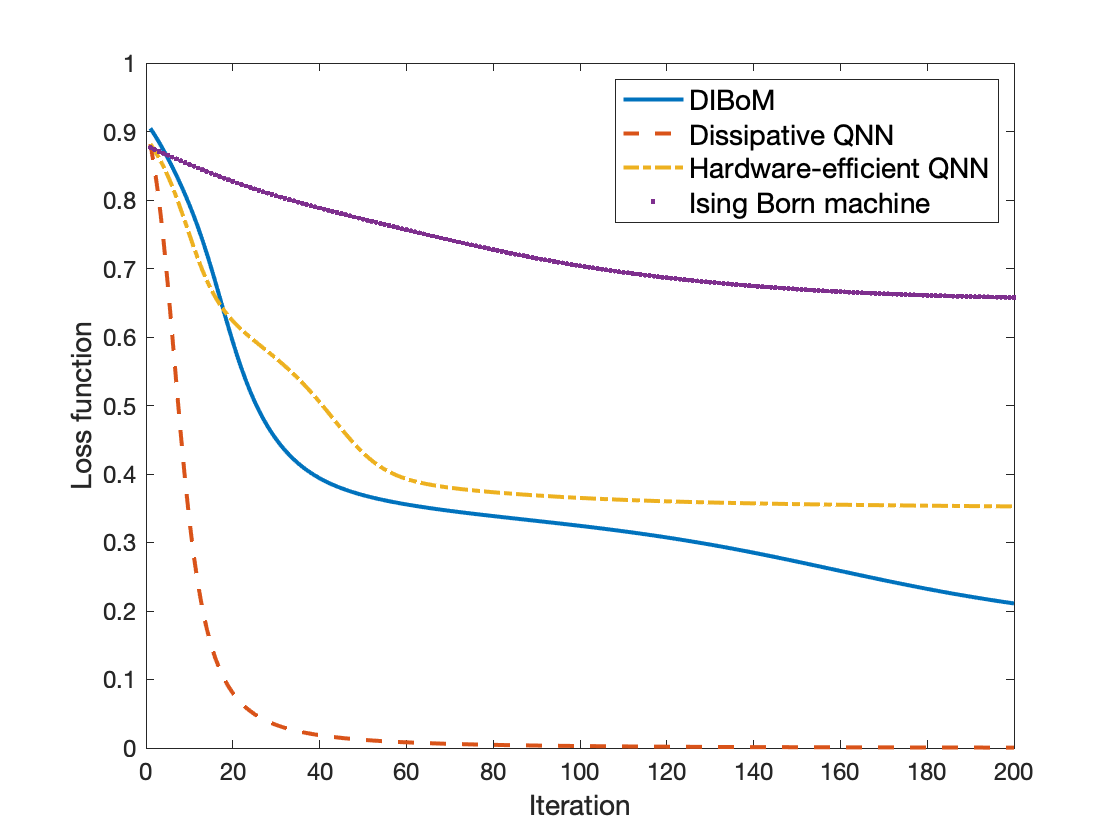}
\caption{Comparison of a DIBoM with a dissipative QNN,  a hardware-efficient QNN and an Ising Born machine in terms of the performance for a random unitary $V$. 
}
\label{fig:baseline_random}
\end{figure}

To facilitate a performance comparison between the DIBoM and other models, we have assumed that the hidden unitary $V$ has the same structure as the DIBoM, but with unknown parameters. To further evaluate the DIBoM's performance, we conduct a test where $V$ is a random unitary, with $n=3$ qubits and $10$ layers for both the DIBoM and hardware-efficient QNN. The results are shown in Fig.~\ref{fig:baseline_random}. The DIBoM outperforms the hardware-efficient QNN and the Ising Born machine but has lower accuracy than the dissipative QNN. As compensation, the number of parameters in a DIBoM scales only quadratically with $n$, whereas the dissipative QNN has exponential scaling.

\subsection{Ablation study}
\label{sec:ablation}

After analyzing the overall performance of the DIBoM architecture, we next 
alter some components of the DIBoM to get a better understanding of the contribution of each component
of the DIBoM to its performance. 

\begin{figure}[htb]
\centering \includegraphics[width=8.5 cm]{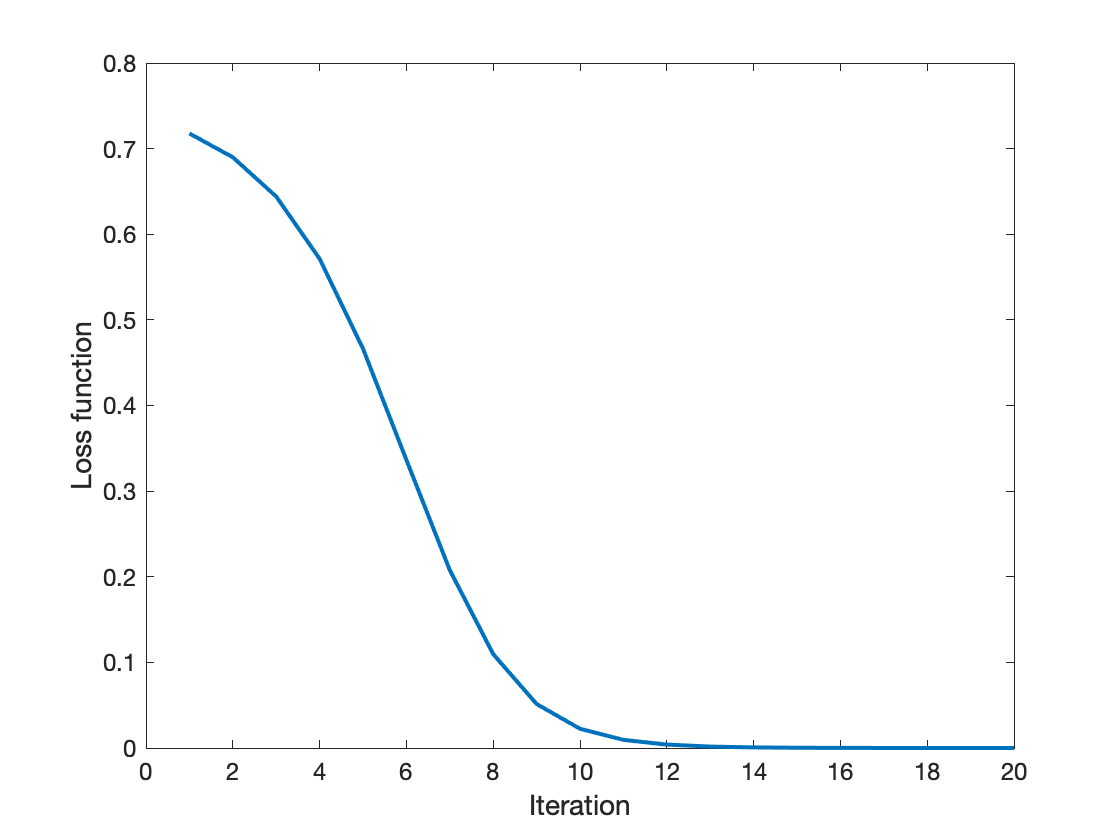}
\caption{Learning curve of a variant of the DIBoM ($U_{SG}^2$). 
}
\label{fig:plotSG}
\end{figure}

\begin{figure}[htb]
\centering \includegraphics[width=8.5 cm]{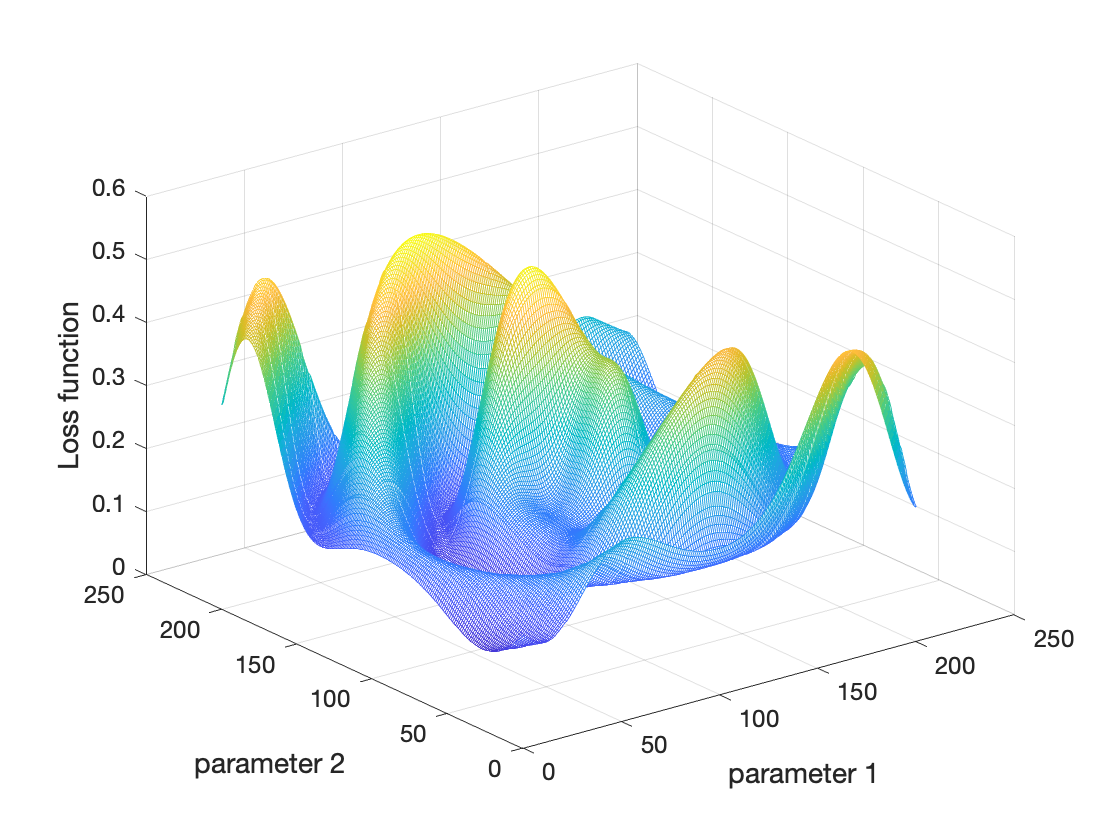}
\caption{Optimization landscape of a variant of the DIBoM ($U_{SG}^2$) with two parameters varying and other parameters fixed.
 }
\label{fig:landscape2}
\end{figure}

First, we investigate the case that the DIBoM consists of a layer of one single-qubit gate ($U_{SG}^2$), which acts exclusively on the second qubit. The training result is plotted in Fig.~\ref{fig:plotSG}, where it is evident that the fidelity of the model output reaches 1 after sufficient training. Moreover, it converges rapidly, reaching zero loss by the 14th iteration. To delve deeper into the optimization process and understand why the optimization of the single-qubit unitary does not trap at a local minimum, we plot the loss function as a function of two parameters while keeping all other parameters fixed. Both parameters are from the single-qubit unitary, and the resulting plot is depicted in Fig.~\ref{fig:landscape2}. Despite the highly non-convex optimization landscape, it is noticeable that all local minima have approximately the same value, explaining the achievement of zero loss during training of a single-qubit unitary.

\begin{figure}[htb]
\centering \includegraphics[width=8.5cm]{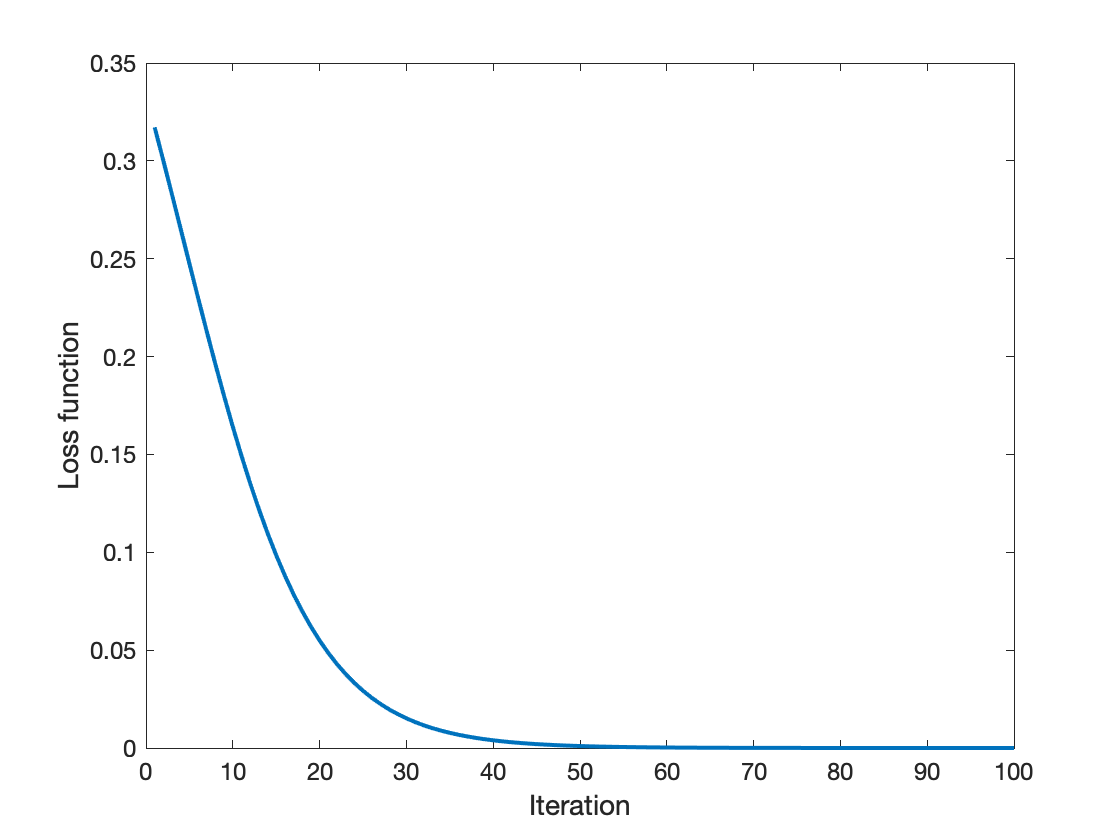}
\caption{Learning curve for a variant of the DIBoM ($U_{CZ}$). 
}
\label{fig:plotCZ}
\end{figure}

Next we examine the case that the DIBoM consists of one layer of generalized CZ gates ($U_{CZ}$). As displayed in Fig.~\ref{fig:plotCZ}, the fidelity of the model output approaches unity after sufficient training, similar to the case of single-qubit gates.  Note however that the initial loss for the generalized CZ gate case is lower than that of the single-qubit gate case. This observation suggests that a generalized CZ gate is more rigid than a single-qubit gate, implying a narrower range of variation in the quantum output induced by the former. Note also that convergence for the generalized CZ gate case occurs around the 50th iteration, which is slower than the convergence rate for the single-qubit gate case, indicating the former case is comparatively harder to train than the latter case.

\begin{figure}[htb]
\centering \includegraphics[width=8.5cm]{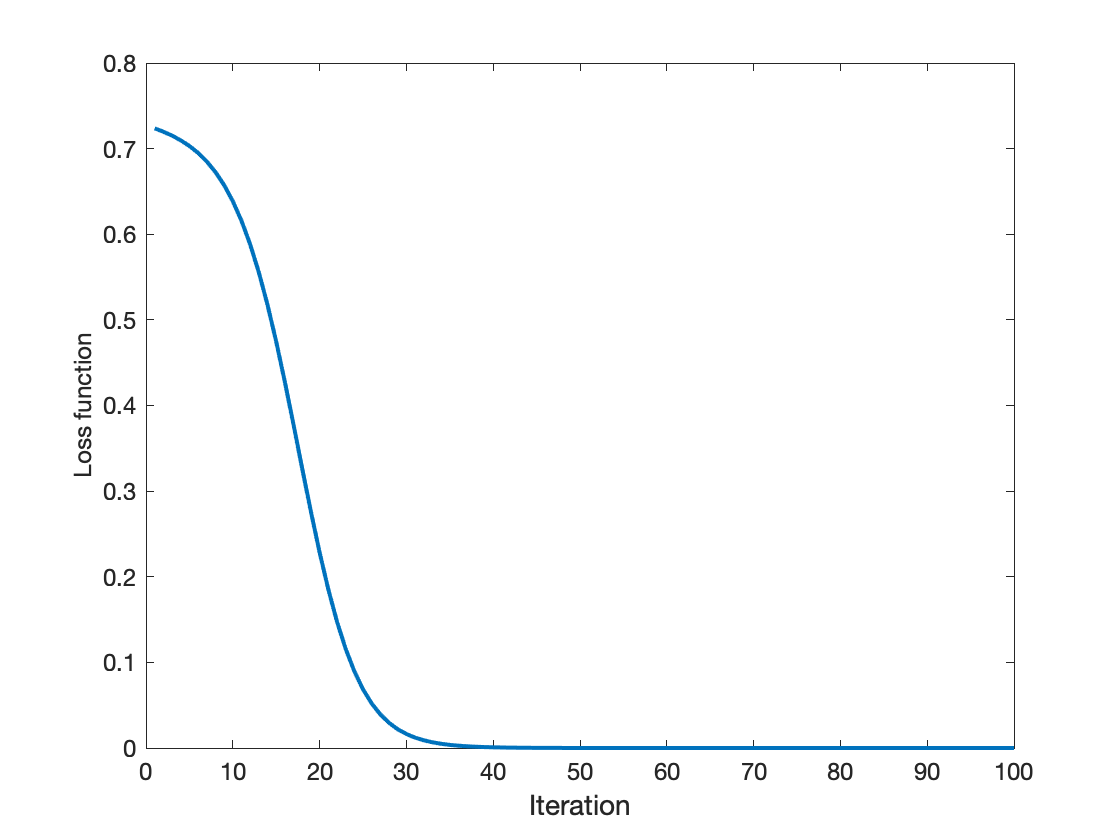}
\caption{Learning curve of a DIBoM ($U_{CZ} U_{SG}^2 $) with the first layer trained and the second layer fixed. 
}
\label{fig:fixedsecond}
\end{figure}

\begin{figure}[htb]
\centering \includegraphics[width=8.5cm]{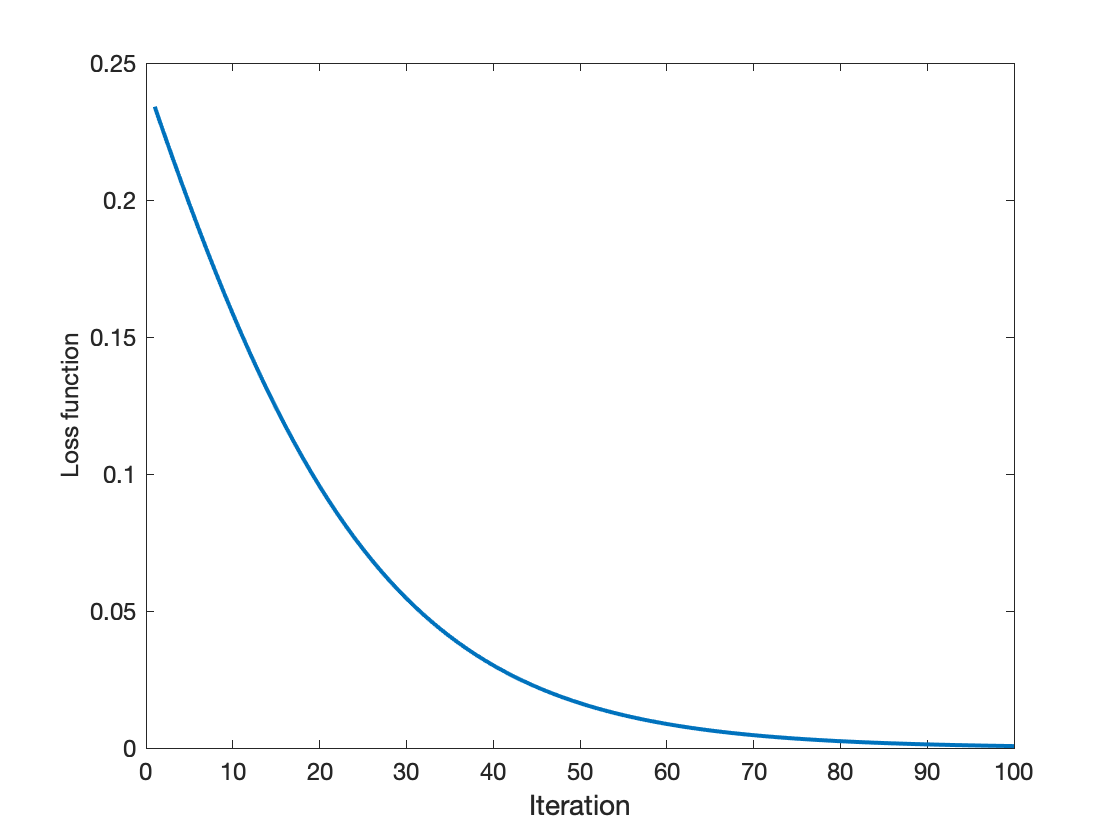}
\caption{Learning curve of a DIBoM ($U_{CZ} U_{SG}^2 $) with the second layer trained and the first layer fixed. 
}
\label{fig:fixedfirst}
\end{figure}

Then we investigate the DIBoM ($U_{CZ} U_{SG}^2 $) where one of the layers is fixed. For the case that the second layer is fixed, the training result is shown in Fig.~\ref{fig:fixedsecond}. As illustrated, the training converges to zero loss and the convergence is quite fast, reaching the optimal loss at the 40th iteration. Conversely, for the case that the first layer of the DIBoM is fixed while the second layer is to be trained, the convergence is slower, as evidenced by the training results shown in Fig.~\ref{fig:fixedfirst}. This slower convergence could be attributed to the flatter optimization landscape of the generalized CZ gate. Nonetheless, the training also eventually converges to zero loss.

\begin{figure}[htb]
\centering \includegraphics[width=8.5cm]{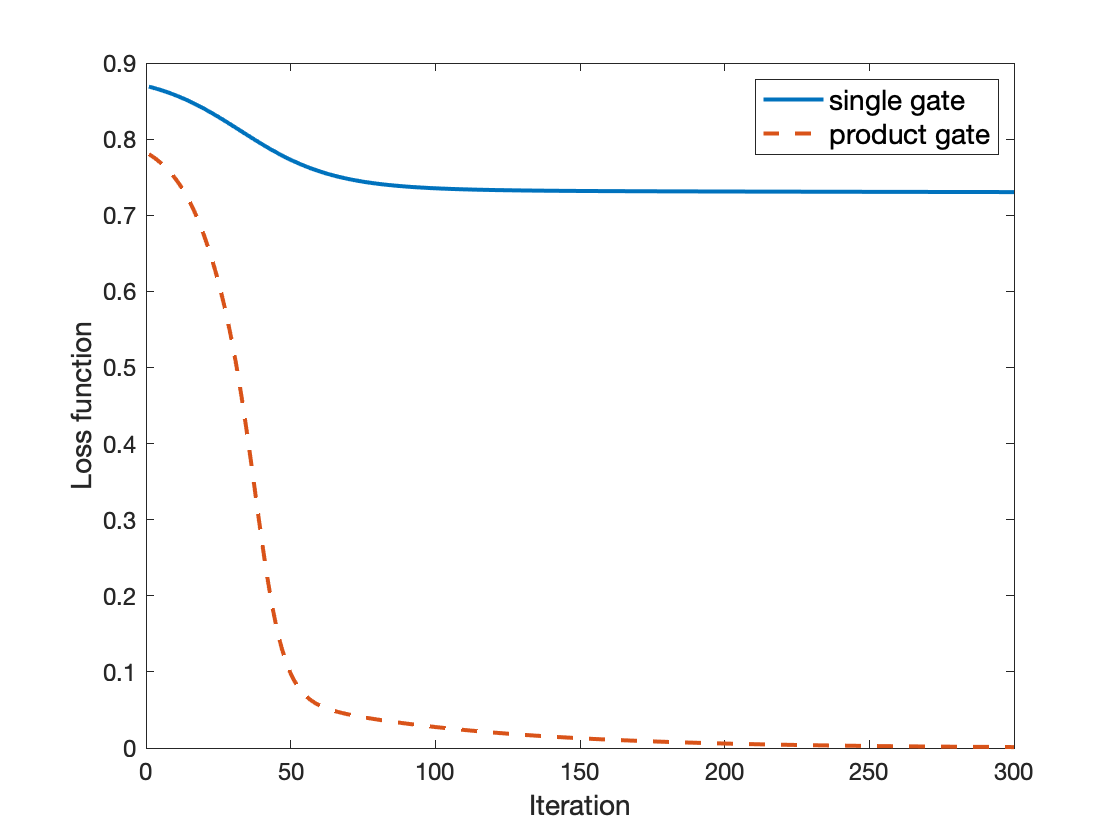}
\caption{The performance comparison between the product-gate variant of the DIBoM  ($U_{CZ} U_{SG}^\otimes $) and a DIBoM ($U_{CZ} U_{SG}^2 $). 
}
\label{fig:product_gate}
\end{figure}

Next we examine a network comprising a layer of single-qubit gates on all qubits followed by a layer of generalized CZ gates ($U_{CZ} U_{SG}^\otimes $). Here, $U_{SG}^\otimes$ is a $n$-qubit gate that can be decomposed as a tensor product of parameterized single-qubit gates, potentially featuring varying parameters.
We refer to the former layer as a \emph{product gate} layer.
Figure~\ref{fig:product_gate} compares the performance of the product gate case to the single gate setting. 
 It can be seen that the product gate case converges to a zero loss while the single gate case does not, implying that the product gate case has more expressive power.

\begin{figure}[htb]
\centering \includegraphics[width=8.5cm]{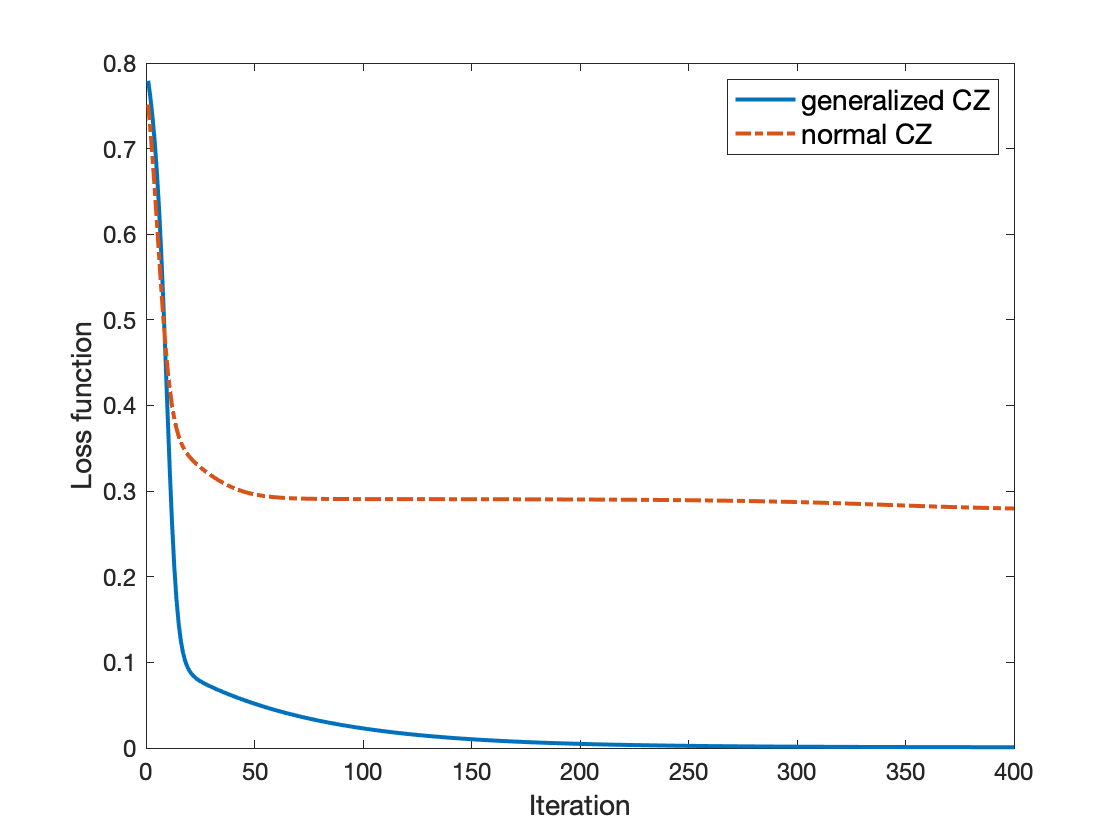}
\caption{Performance comparison between a DIBoM and an ablated DIBoM where generalized CZ gates are replaced by normal CZ gates. }
\label{fig:ab_GCZ}
\end{figure}

An ablation study of generalized CZ gates is then conducted by comparing the following two models. The first model, denoted as ``generalized CZ'', utilizes DIBoM with 3 layers ($U_{SG}^\otimes U_{CZ} U_{SG}^\otimes$). The second model, denoted as ``normal CZ'', is almost identical to the first model, except that all the generalized CZ gates in the second layer are replaced with normal CZ gates. The study's results are presented in Fig.~\ref{fig:ab_GCZ}, which indicates that the first model achieves a much lower loss than the second model. This finding suggests the strong effectiveness of generalized CZ gates.

\begin{figure}[htb]
\centering \includegraphics[width=8.5cm]{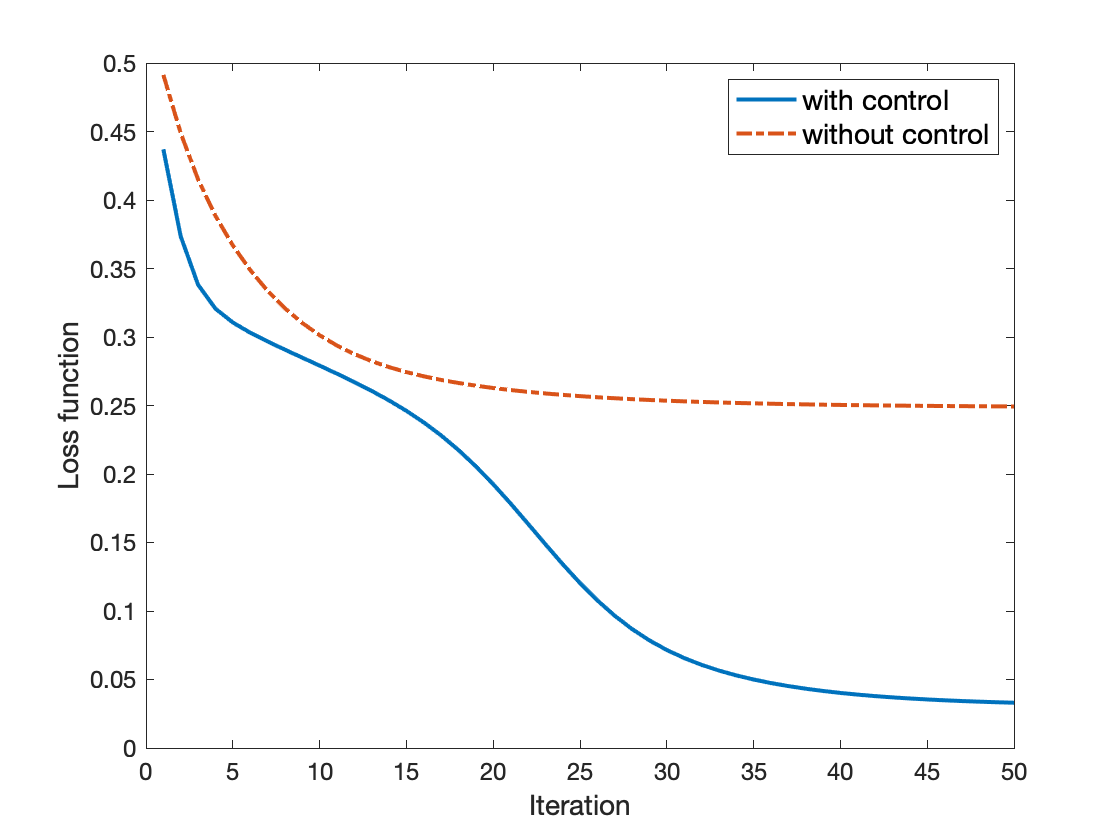}
\caption{The performance comparison between a full DIBoM with controlled gates and an ablated DIBoM without controlled gates. }
\label{fig:abcontrol}
\end{figure}

To evaluate the effect of the controlled unitary $V_j$, we conduct an ablation study. For this purpose, we modify the training dataset to have different input and output dimensions. Specifically, we construct the training dataset as ${( \ket{\psi_i}\otimes \ket{0} \otimes \ket{0} , \ket{\psi_i})}_{1\le i \le N}$, where $\ket{\psi_i}$ represents a randomly generated pure qubit using the Haar measure. The objective of the model is to transform the quantum sample such that the third qubit approximates the quantum label as closely as possible.
We denote the complete DIBoM model as \emph{with control}. The model first applies a three-qubit unitary $U$, measures the first two qubits in the computational basis, and then uses the measurement result $j$ to perform the controlled unitary $V_j$ on the third qubit. On the other hand, the ablated version, denoted as \emph{without control}, includes the same first two components but lacks the final component. More precisely, the unitary $U$ is composed of three layers $U_{SG}U_{CZ}U_{SG}$, and there are four single-qubit unitaries $V_j$ ($0\le j \le 3$) to be optimized.
The results are illustrated in Fig.~\ref{fig:abcontrol}, which indicates that the complete DIBoM model achieves significantly lower loss than the ablated version. This suggests that the complete DIBoM model can effectively learn the quantum teleportation protocol from scratch, while the ablated version lacks this capability.

\subsection{Parameter sensitivity}
\label{sec:sensitivity}

So far, we have investigated the performance of the DIBoM and the contributions of its individual components to this performance. For a more comprehensive evaluation of the DIBoM, we conduct experiments to test the effect of different parameters, including the data size, the number of layers, the number of qubits per layer, and the regularization constant.

\begin{figure}[htb]
\centering \includegraphics[width=8.5 cm]{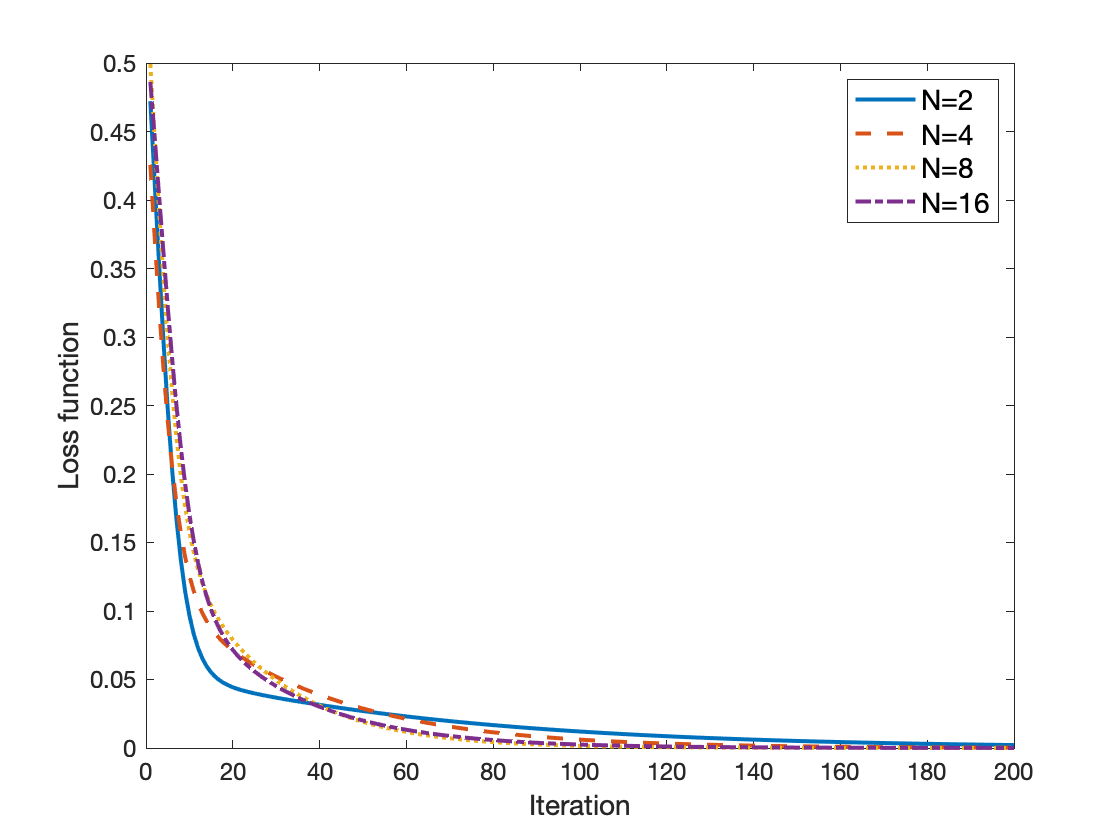}
\caption{The training performance of a DIBoM with a varying number $N$ of training samples. 
}
\label{fig:sampleTraining}
\end{figure}

\begin{figure}[htb]
\centering \includegraphics[width=8.5cm]{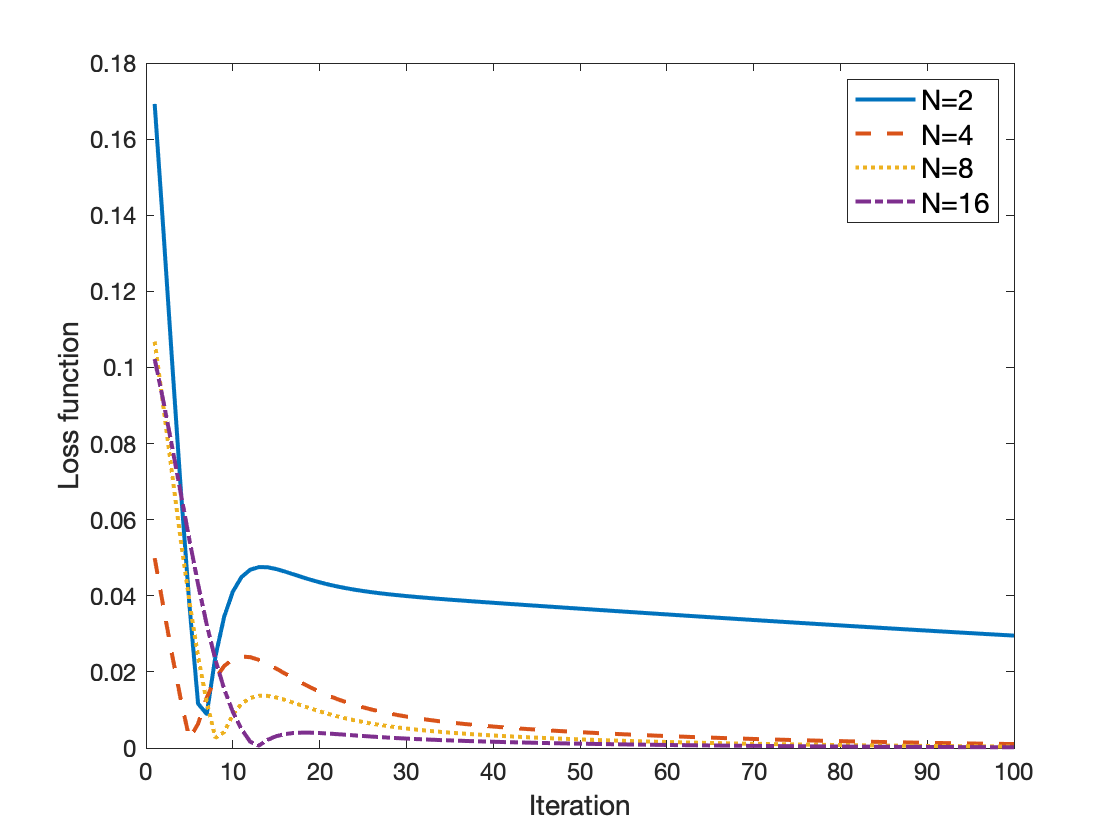}
\caption{The gap between the training and test performance of a DIBoM with  $N$ training samples. 
}
\label{fig:sampleGap}
\end{figure}

We begin by examining the relation between the number of training samples $N$ and the performance characterized by the training loss. 
The result is shown in Fig.~\ref{fig:sampleTraining}, which clearly demonstrates that the larger the sample size, the 
faster the convergence. This is because more samples
help to estimate the intrinsic unitary $V$ better during training. 
We next test the relation between the number of samples and the gap between
the training performance and the test performance, as shown in Fig.~\ref{fig:sampleGap}. It can be seen that
 as the number of samples increases, the gap gradually becomes smaller.
 This is expected as a larger number of samples results in a smaller variance and thus, improved generalization performance.

\begin{figure}[htb]
\centering \includegraphics[width=8.5cm]{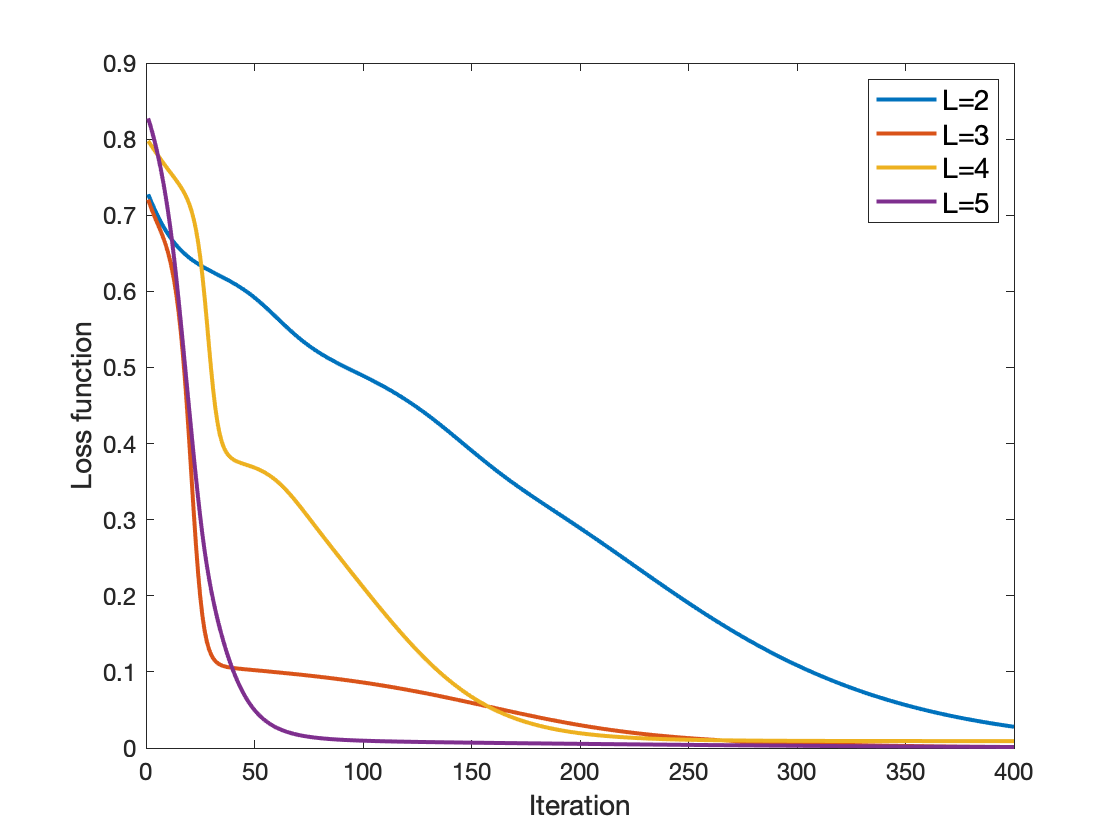}
\caption{The performance of a DIBoM with a different number $L$ of layers. 
}
\label{fig:variousLayer}
\end{figure}

Next, we test the effects of the layer number $L$ 
on the performance of the product gate variant of the DIBoM ($\cdots U_{SG}^\otimes U_{CZ} U_{SG}^\otimes $), as shown in Fig.~\ref{fig:variousLayer}. It can be seen that 
the losses of all cases converge to zero loss, indicating that the network can scale to many layers without adversely affecting trainablity.

\begin{figure}[htb]
\centering \includegraphics[width=8.5 cm]{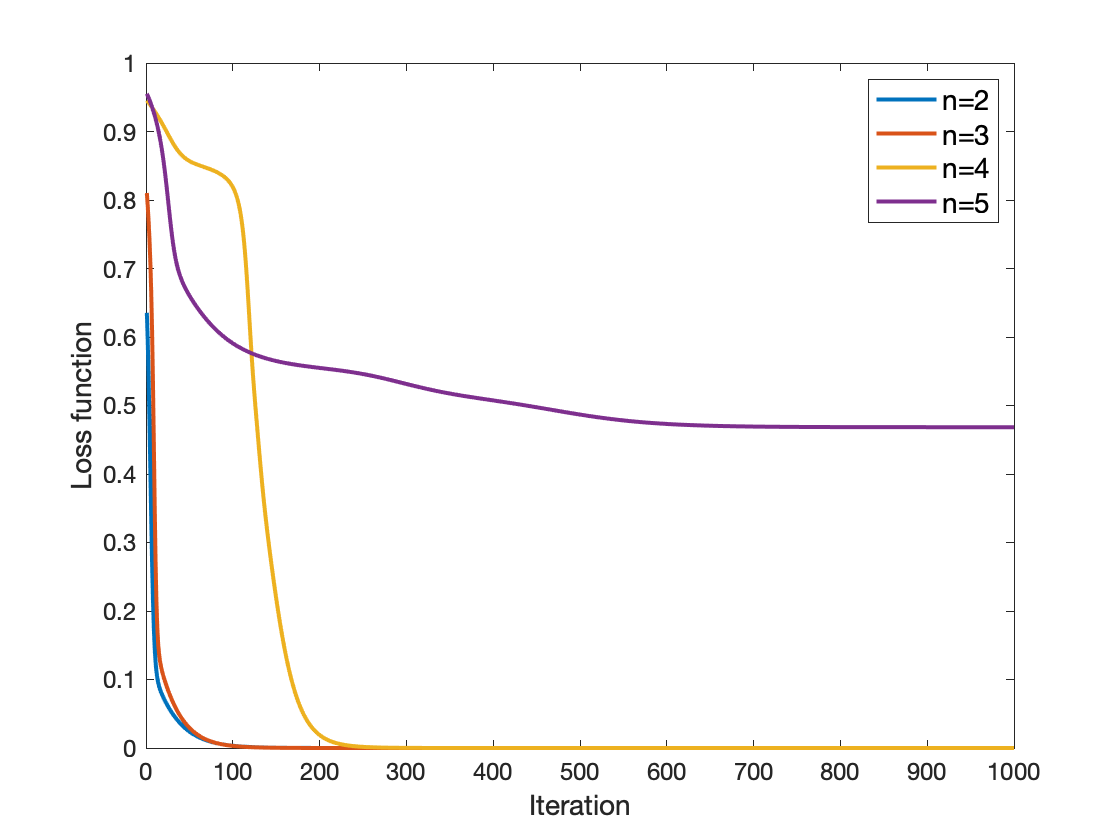}
\caption{The performance of a DIBoM with a different number of qubits $n$ per layer. 
}
\label{fig:variousqubit}
\end{figure}

\begin{figure}[htb]
\centering \includegraphics[width=8.5 cm]{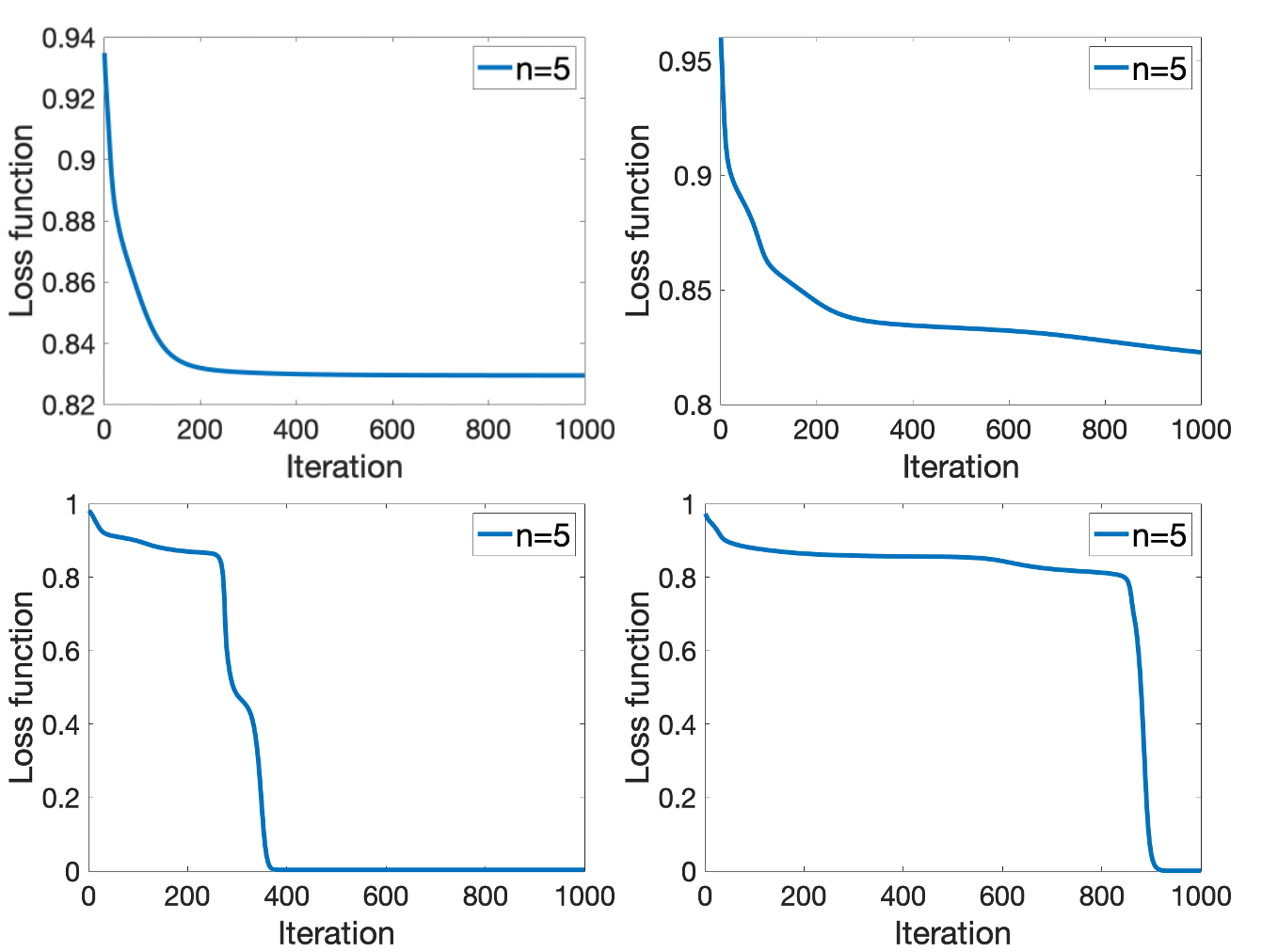}
\caption{Four learning curves of a DIBoM with 5 qubits per layer. 
}
\label{fig:n=5}
\end{figure}

Then we test the number of qubits $n$ 
that the input and output contain, also on the product gate variant of the DIBoM ($ U_{CZ} U_{SG}^\otimes$).
 As shown in Fig.~\ref{fig:variousqubit}, 
for all cases from $2$ qubits to $4$ qubits, the DIBoM can train well, with zero loss convergence.
However, the case of $n=5$ did not perform as well, resulting in non-zero loss. Therefore, we perform four additional simulations for $n=5$, which are displayed in Fig.~\ref{fig:n=5}.
It is evident that although the loss function eventually converges to zero loss, this requires a larger number of iterations. Moreover, the training curve displays both slow-varying and fast-varying regions, a phenomenon that is already observable in the case of $n=4$, but is more pronounced when $n=5$.

\begin{figure}[htb]
\centering \includegraphics[width=8.5 cm]{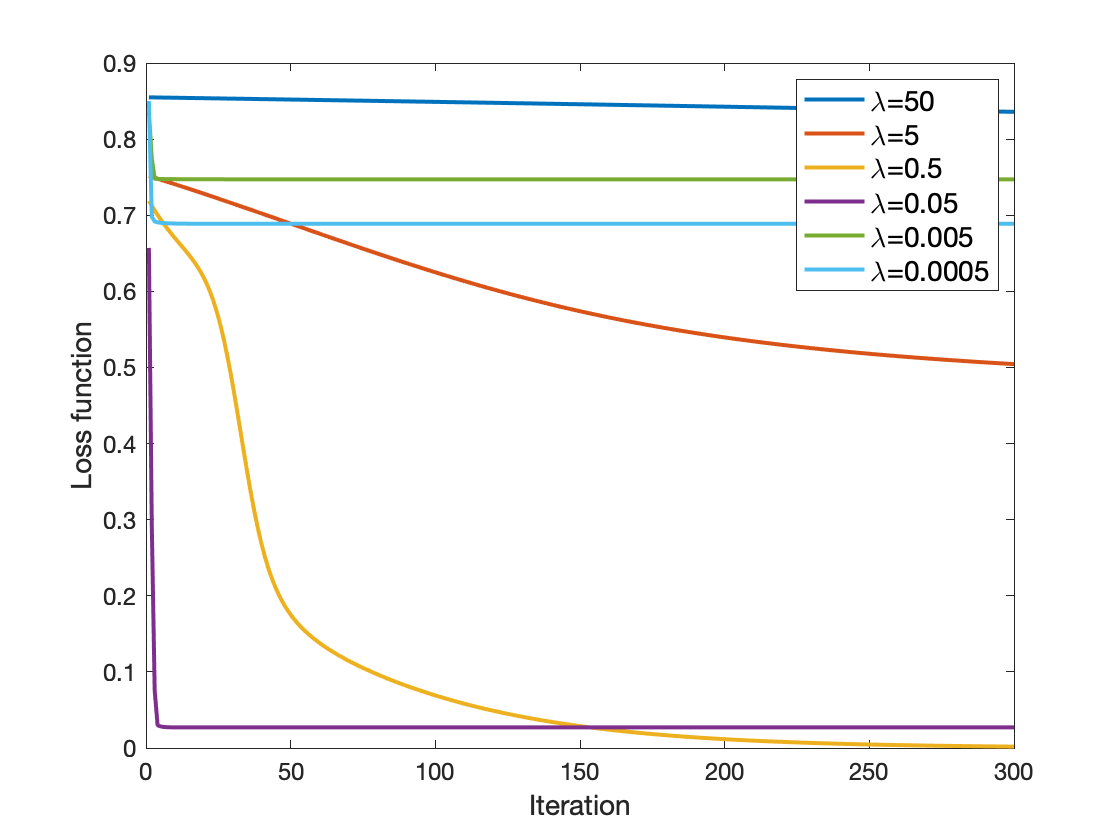}
\caption{The performance of a DIBoM with different parameters of the regularity constraint  $\lambda$. }
\label{fig:variouslambda}
\end{figure}

In addition, we  test the effect of the regularity constraint parameter $\lambda$ 
on the performance of  the DIBoM. 
The results, shown in Fig.~\ref{fig:variouslambda}, indicate 
an optimal value of $\lambda$ at 0.5. Deviating from this value leads to worse convergence.

\begin{figure}[htb]
\centering \includegraphics[width=8.5cm]{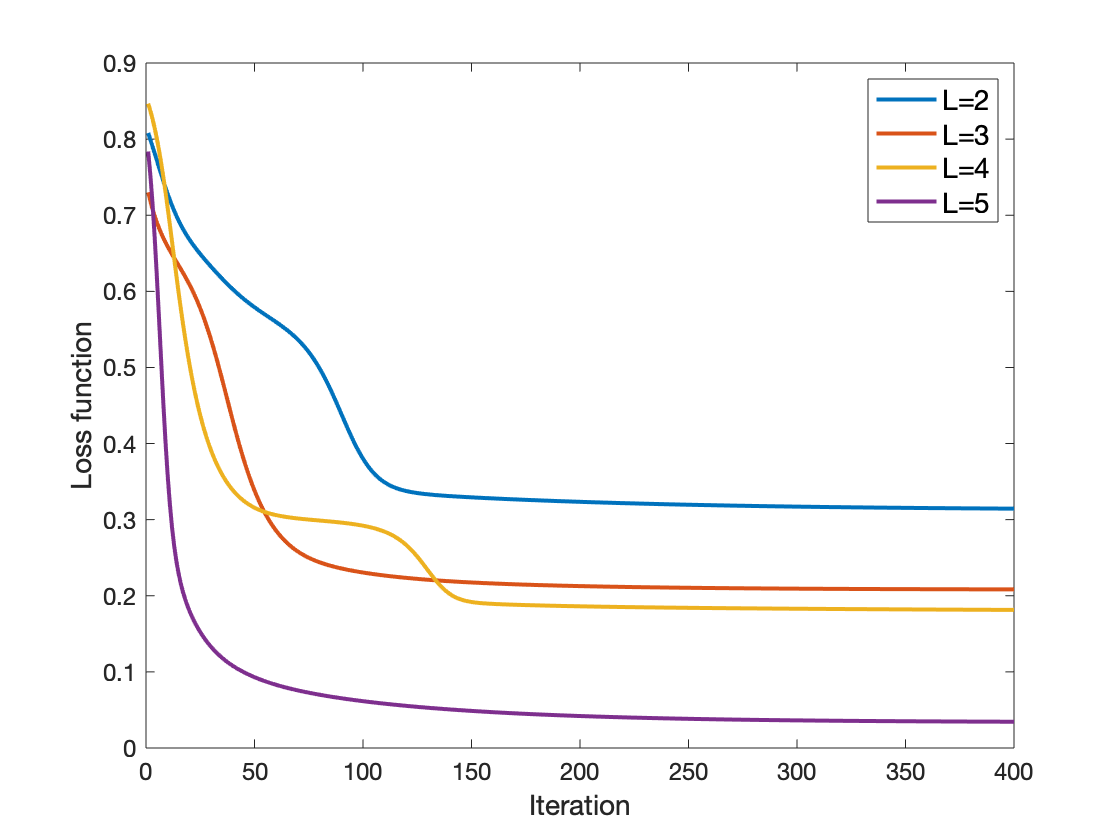}
\caption{The performance of a DIBoM with $L$ layers on a generic quantum learning dataset. 
}
\label{fig:full_unitary}
\end{figure}

Finally, we assess the ability of a DIBoM ($\cdots U_{SG}^\otimes U_{CZ} U_{SG}^\otimes $) to approximate an arbitrary unitary using 2 to 5 layers, alternating between product gate and generalized CZ gate layers. The results, presented in Fig.~\ref{fig:full_unitary}, demonstrate that the converged loss decreases as the number of layers increases. Notably, with only 5 layers, the DIBoM already achieves a low loss.

\subsection{Robustness to noise}
\label{sec:robustness}

\begin{figure}[htb]
\centering \includegraphics[width=8.5 cm]{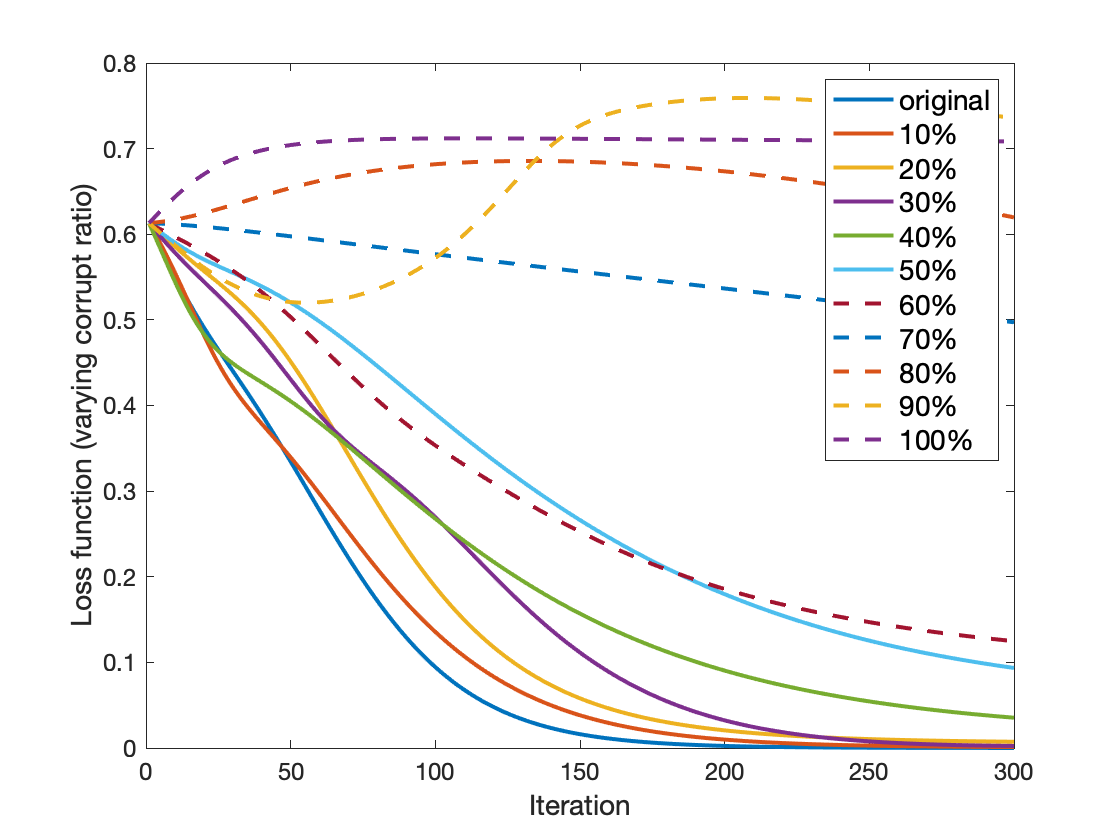}
\caption{The performance of a DIBoM with various proportions of the training data corrupted. 
}
\label{fig:corrupt}
\end{figure}

We further assess the 
robustness of a DIBoM to noise in the data since real-world data inevitably contains noise. To this end, we manually corrupt 
some of the training data and evaluate its effect on the test loss. The corruption ratio of the training data varies from $10\%$ to $100\%$, with an uncorrupted dataset, denoted by ``original'', acting as the control. Given a corruption ratio, say $30\%$, we randomly select $30\%$ of real quantum data and replace it with fake data 
$( \ket{ \phi^{in}_x},\ket{\phi^{out}_x})$, where $\ket{ \phi^{in}_x}$ and $\ket{\phi^{out}_x}$ are Haar random $n$-qubit pure states. 
The results are visualized in Fig.~\ref{fig:corrupt}.
On the negative side, with a gradual increase of the corrupted ratio, the performance gradually degrades, as evidenced by the comparison of solid and dashed lines. On the positive side, even with up to $60\%$ of corrupted data, the DIBoM remained effective, which indicates a high level of robustness against noise.

\begin{figure}[htb]
\centering \includegraphics[width=8.5 cm]{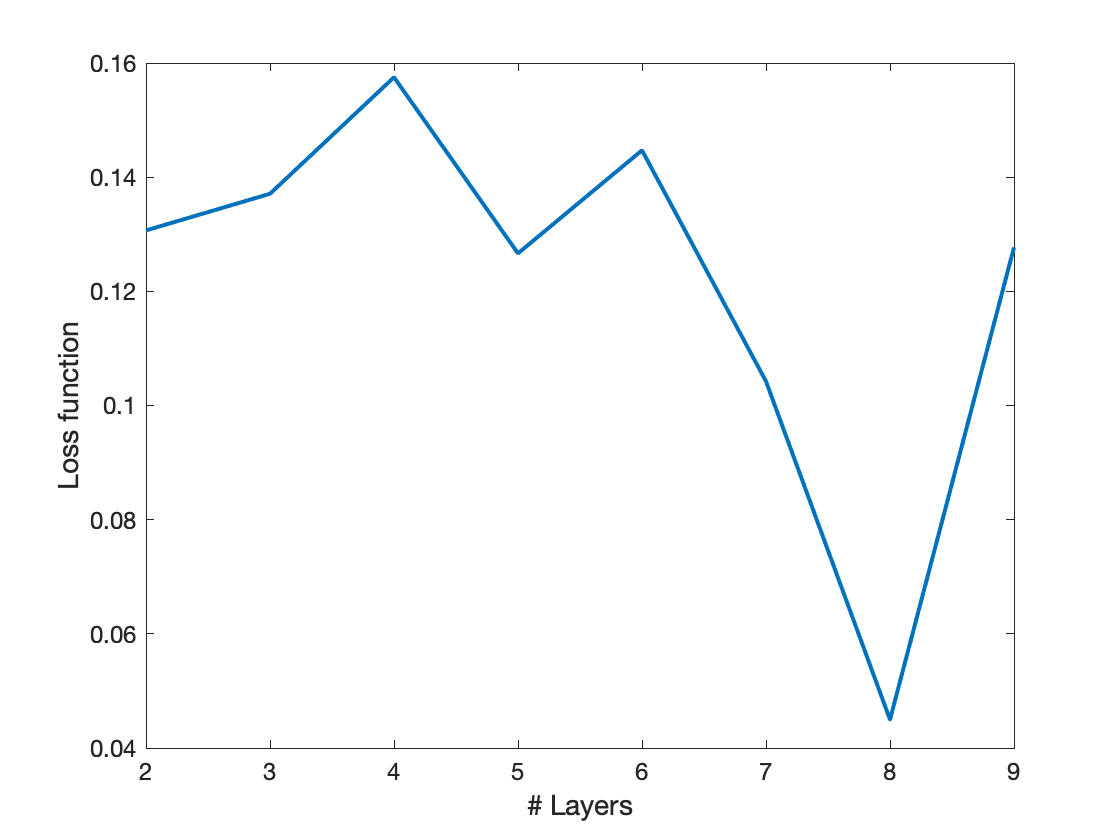}
\caption{Noise robustness of the DIBoM with different numbers of layers. }
\label{fig:robustVSlayers}
\end{figure}

We also investigate the noise robustness of the model with respect to varying numbers of layers. 
To perform this investigation, we set a fixed corruption ratio of $20\%$ and vary the number of layers, while monitoring the loss at the 300 iterations. The results are presented in Fig.~\ref{fig:robustVSlayers}. Notably, our analysis reveals that the noise robustness of the model remains consistent across different numbers of layers. This finding highlights the scalability of the DIBoM and suggests that increasing the number of layers does not negatively impact the noise robustness of the model.

\subsection{Barren plateau}
\label{sec:barren}

\begin{figure}[htb]
\centering \includegraphics[width=6cm]{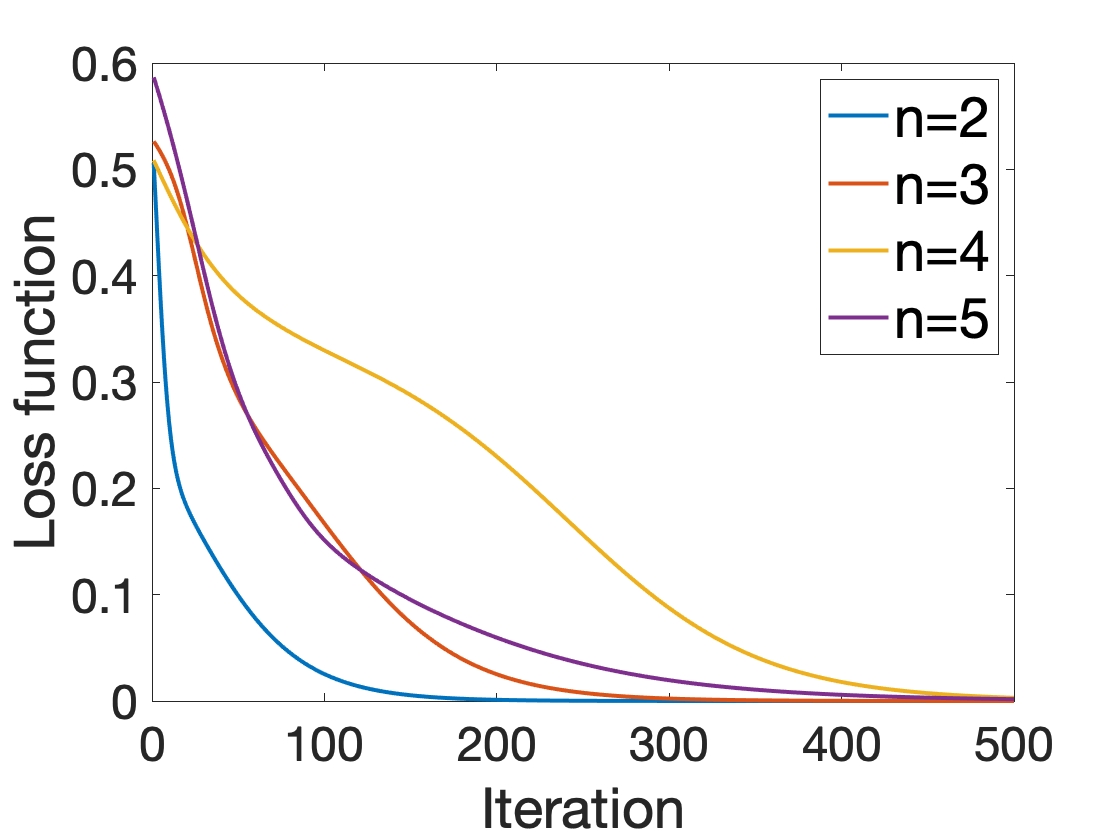}
\caption{The performance of a DIBoM with a local cost function. Here, $n$ is the number of qubits.
}
\label{fig:localcost}
\end{figure}

In Sec.~\ref{sec:sensitivity}, we have observed that the model with the global loss function Eq.~\eqref{eq:lossfunction} already suffers from the barren plateau (slow-varying region) issues for $n=5$ qubits.
Previous work by Cerezo et al. \cite{cerezo2021cost} demonstrated that local cost functions can mitigate this issue.
Hence we incorporate the local cost function described in Eq.~\eqref{eq:locallossfunction}.
We defer the details of the classical simulation of the model under the local loss function to Appendix~\ref{appsec:localcost}.
To accommodate the local cost function, we design the training data as a product state $\ket{\phi_{in}}=\ket{\phi_{in}}_1 \otimes \dots \otimes \ket{\phi_{in}}_n$ for $n$ qubits, which we refer to as \emph{product-form training data}.
Notably, a zero local cost function for this data implies a zero global cost function. 
We examine the performance of the local cost function by training the model with 2 to 5 qubits, as shown in Fig.~\ref{fig:localcost}. We observe that  all learning curves converge to zero loss quickly, suggesting that the barren plateau issue is mitigated. This stands in stark contrast to the global cost function, which exhibits the barren plateau phenomenon when $n=5$.

\begin{figure}[htb]
\centering \includegraphics[width=8.5cm]{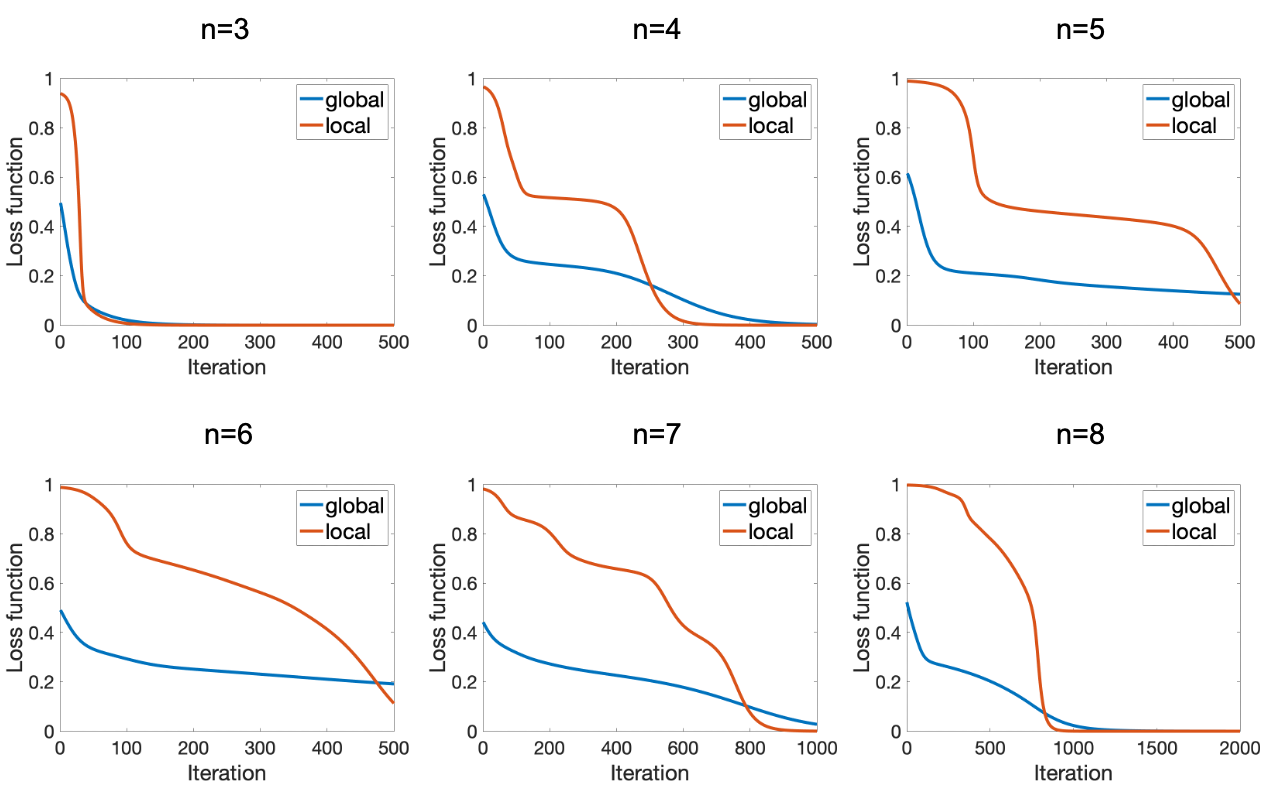}
\caption{Comparisons between global and local cost functions with different numbers of qubits $n$. Here, $n$ ranges from 3 to 8. The redline depicts the local cost function while the blue line depicts the global cost function.
}
\label{fig:globalVSlocal}
\end{figure}

We have also plotted the comparison between local and global cost functions in Fig.~\ref{fig:globalVSlocal}.
The local cost function always reaches zero loss faster than the global cost function. More 
importantly, the curve of the local cost function lacks a flat region,
suggesting that the barren plateau phenomenon is mitigated for the DIBoM with a local cost function.
Notably, we observe that the barren plateau issue is also mitigated for a global cost function with product-form training data. This observation suggests that product-form training data may be easier to train than entangled training data.

 Finally, we discuss the computation time required for our simulations.
 For the case $n=8$, which is the largest simulation we performed, the computation takes about 1 hour on an 8-core 3.2GHz computer.
The number of parameters for this case is $4n + n(n-1)/2=60$.
It is worth noting that the
 computation time is exponential in $n$, which is a consequence of the classical computation requiring multiple multiplications on density matrices of size $2^n \times 2^n$ in each iteration.
Each of these multiplications
 takes time $2^{3n}$, resulting in a cost of $256^3$ per multiplication when $n=8$. Furthermore, the number of iterations is polynomially
 related to $n$, with approximately 2000 iterations required for $n=8$. Hence, the computation burden on a classical computer is substantial. However, it is important to note that intermediate-scale quantum computers have the potential to substantially decrease computation time to a polynomial of $n$, as the classical manipulation of $2^n \times 2^n$ density matrices is no longer necessary. As such, we expect that current noisy intermediate-scale quantum (NISQ) devices will boost the trainable size to a few hundred qubits.

\section{Discussion}
\label{sec:discussion}

In summary, we examined a deep Ising Born machine (DIBoM) and showed it has a good balance between efficiency and universality. Specifically, we described its model architecture and its training procedure. Additionally, through theoretical analysis, we demonstrated that the DIBoM has the capability of universal quantum computation. Apart from the theoretical analysis, we empirically evaluated the performance of the DIBoM and compared it with other QNNs. Our evaluations revealed that the DIBoM has a moderate number of parameters while being quite expressive. Along the way, we introduced a new expressivity measure called fidelity-based expressivity, which may be of independent interest.

There are two potential limitations of the DIBoM: trainability and generalizability. Trainability refers to the ability to find the global minimum in a polynomial number of iterations with respect to the number of qubits. In our simulations, we have shown that the DIBoM is trainable for a moderate number of qubits, but it is unclear whether it remains trainable for a large number of qubits, which is beyond the capability of our simulation. Moreover, there is no theoretical guarantee that the DIBoM is trainable, and recent negative results \cite{anschuetz2022quantum} suggest that most shallow and local QNN architectures are not trainable.

Generalizability refers to the ability to achieve low test error given a small training error. In our simulations, we have empirically observed that the DIBoM has low test error, but we have no theoretical guarantee for this fact. When the number of parameters of a QNN is much larger than the number of training data, over-parameterization can cause overfitting, making it difficult to achieve theoretical guarantees of generalizability. This is a challenging problem even for classical neural models.

Therefore, it is crucial to develop QNN architectures that achieve efficiency, universality, provable trainability, and provable generalizability simultaneously. The DIBoM only addresses the first two goals, leaving much room for improvement.

There are a few other promising avenues for future research. First, applying the DIBoM to downstream quantum learning tasks is likely to be both fruitful and interesting. Second, an experimental demonstration of the DIBoM on quantum hardware would be interesting. Third, due to the exponential cost of the classical simulation, a NISQ device may show a speed advantage in training the DIBoM, which makes it an ideal target for showing quantum supremacy on practical problems.

The tradeoff between efficiency and universality is also worth further investigation. To achieve universality as defined in our work, an ansatz needs to have exponentially many parameters because the ansatz cannot express all unitaries if the dimension of its Hilbert space is smaller than that of $\mathsf{SU}(n)$. This implies full universality and efficiency cannot be achieved simultaneously for any quantum learning model. There are several directions to further explore the tradeoff between universality and efficiency and the associated design of quantum learning models.

First, by relaxing the definition of universality, there may exist more interesting tradeoff between universality and efficiency. However, this makes the research landscape more complex since there are a lot of ways to weaken the notion of universality. Previous research has considered weakening the universality to the class of functions that maps 0 to the ground state of a Hamiltonian which is the sum of poly($n$) Pauli bases \cite{biamonte2021universal}, that is real and continuous \cite{goto2021universal}, that can be described by a quantum circuit with a polynomial number of gates where each gate acts on a constant number of qubits \cite{cai2022sample}, and that is boolean \cite{herman2022expressivity}. Besides these choices, there are many other choices available, potentially infinitely many. For example, one can consider the class of function that maps 0 to the thermal state of a Hamiltonian which is the sum of poly($n$) Pauli bases, that is complex and meromorphic, that can be described by a quantum circuit with a logarithmic number of gates where each gate acts on a logarithmic number of qubits, to just name a few. How to achieve these different types of universality while maintaining efficiency in a strict sense is an interesting research question.

Another direction is to replace universality by restricting the ansatz to contain the solution one is looking for. In this case, the ansatz becomes problem specific, which is not a universal ansatz that can deal with all learning problems. Following this line, after taking a learning problem, one should design a specific ansatz that suits this problem which requires additional manual work and expertise. How to reduce the manual efforts in designing a specific ansatz for a given learning problem (such as combinatorial optimization problems \cite{zhou2020quantum}, learning the ground state of a Hamiltonian \cite{motta2020determining}, simulating quantum dynamics \cite{sparrow2018simulating}, or drug discovery for a specific disease \cite{cao2018potential}) is an interesting question on its own.

\begin{acknowledgements}
This work was supported by the National Natural Science Foundation of China (12105105), the Natural Science Foundation of Shanghai (21ZR1415800), the Shanghai Sailing Program (21YF1409800), and the startup fund from East China University of Science and Technology (JKH01221665 and YH0142206).
\end{acknowledgements}

\appendix

\section{Evaluation details of the Fidelity-based Measure}
\label{appsec:fbe}

Here, we detail how we numerically evaluate an upper bound of the fidelity-based measure (FBE) for any QNN architecture in Fig.~\ref{fig:expressivity}.

For a QNN architecture $\mathsf{A}(\theta)$, we first select $k=100$ random unitaries $U_i$ ($1 \le i \le k$) and $m=10$ random pure states $\phi_j$  ($1 \le j \le m$).
Then for any $U_i$, we optimize $\theta$ to minimize 
\begin{equation}
s_i =  \frac{1}{m} \sum\limits_{j=1}^{m}  | \bra{\phi_j}  \mathsf{A}(\theta)^\dagger U_i \ket{\phi_j}|.
\end{equation}
The bound $\min_i s_i $ is then taken as the upper bound of the FBE.

\section{Concrete number of parameters for a universal 3-qubit DIBoM}
\label{appsec:3qubit}

In this section, we provide the precise number of parameters needed for a 3-qubit DIBoM to be universal.

To begin with, according to Section 4.5.1 of Ref.~\cite{nielsen2010quantum}, a 3-qubit unitary can be decomposed as a product of $2^{3-1}(2^3-1)=28$ two-level unitaries. Our next goal is to decompose a two-level 3-qubit unitary further.

As per Section 4.5.2 of Ref.~\cite{nielsen2010quantum}, a two-level 3-qubit unitary can be decomposed as a product of at most 5 controlled-controlled single-qubit unitaries. Our next goal is to decompose a controlled-controlled single-qubit unitary further.

Figure 4.18 of Ref.~\cite{nielsen2010quantum} reveals that a controlled-controlled single-qubit unitary can be decomposed as a product of 2 CZ gates and 3 controlled single-qubit unitary gates. Our next goal is to decompose a controlled single-qubit unitary further.

Figure 4.6 of Ref.~\cite{nielsen2010quantum} states that a controlled single-qubit unitary can be decomposed as a product of two CZ gates and single-qubit gates. Consequently, a controlled-controlled single-qubit unitary can be decomposed as a product of $2+3\times 2=8$ CZ gates and single-qubit gates, a two-level 3-qubit unitary as a product of $8 \times 5=40$ CZ gates and single-qubit gates, and a 3-qubit unitary as a product of $40 \times 28=1120$ CZ gates and single-qubit gates. This implies that a $(1120 \times 2+1)$-layer 3-qubit DIBoM is sufficient to realize any 3-qubit unitary.

A $(1120 \times 2+1)$-layer 3-qubit DIBoM contains 1121 single-qubit gate layers and 1120 generalized CZ gate layers. This translates to a total of $1121 \times 9 + 1120 \times 3 = 13449$ parameters.

\section{Derivation of Eq.~\eqref{eq:derivative}}
\label{appsec:derivation}
Expanding the term $dC/ds$ yields
\begin{widetext}
\begin{eqnarray}
\frac{dC}{ds} & = & \lim\limits_{\epsilon \to 0} \frac{ C(s+\epsilon)-C(s)}{ \epsilon} = \lim\limits_{\epsilon \to 0} \frac{  \frac{1}{N} (\sum \limits_{x=1}^N \bra{\phi_x^L} (\rho_x^L(s+\epsilon)- \rho_x^L(s)  ) \ket{\phi_x^L}) }{ \epsilon} \nonumber \\ 
 & = & \frac{1}{N} \sum\limits_x \mathrm{tr}(  \ket{\phi_x^L}\bra{\phi_x^L}
 [ i K^L(s), \prod_{l=L}^1 U^l (s)   
 \rho_x^0  
 \prod_{l=1}^L {U^{l}}^\dagger (s) ] 
 +\dots +  \ket{\phi_x^L}\bra{\phi_x^L} \prod_{l=L}^2 U^l(s) 
 [ i K^{1}(s), U^{1} (s)   \rho_x^0 
 {U^{1}}^\dagger (s) ]  \prod_{l=1}^L {U^{l}}^\dagger (s) ) \nonumber \\
 & = & \frac{1}{N} \sum\limits_x \mathrm{tr}(
 [ \prod_{l=L}^1 U^l (s)  \rho_x^0   \prod_{l=1}^L {U^{l}}^\dagger (s),  \ket{\phi_x^L}\bra{\phi_x^L}]
  i K^L(s) + \dots + [ U^{1} (s)  \rho_x^0  {U^{1}}^\dagger (s) , 
\prod_{l=2}^L {U^{l}}^\dagger (s)  (\ket{\phi_x^L}\bra{\phi_x^L})
 \prod_{l=L}^2 U^l (s) ]
 i K^{1}(s) )  \nonumber   \\
 & = &\frac{i}{N} \sum\limits_x \mathrm{tr} ( \sum_{l=L}^1  M^{l}(s) K^{l}(s) ), 
 \label{eq:derivative31}
\end{eqnarray}
\end{widetext}
where $M^{l}(s)$ is defined as
\begin{eqnarray*}
M^{l}(s)& =&  [  \prod_{j=l}^1 U^{j}(s)  \rho_x^0   \prod_{j=1}^l {U^{j}}^\dagger (s) ,    \\
 && \prod_{j=l+1}^L{U^{j}}^\dagger (s)   \ket{\phi_x^L}\bra{\phi_x^L} \prod_{j=L}^{l+1} U^{j} (s) ].
\end{eqnarray*}
Here $[\cdot,\cdot]$ denotes the commutator operator, and 
the fourth equality in Eq.~\eqref{eq:derivative31} has exploited the relation $\mathrm{tr}(A[B, C]) = \mathrm{tr}([C,A]B)$.

\section{Simulation of the controlled unitary $V_j$}
\label{app:simulate_control}

In this section, we present the classical simulation that involves the controlled unitaries $V_j$. 
Before any measurements, the initial quantum state $\rho^0$ is evolved to the following quantum state: 
\begin{equation}
\rho^1 = U^k \cdots U^1 \rho^0 {U^1}^\dagger \cdots {U^k}^\dagger.
\end{equation}
After measuring with outcome $i$, the post-measurement state is given by
\begin{equation}
\rho^2_i = \mathrm{tr}_A (\rho^1 (\ket{i}\bra{i}_A \otimes I_B) ).          
\end{equation}
The post-measurement states then undergo another series of unitaries and become
\begin{equation}
\rho^3 = \sum_i  V_i^l \cdots V_i^1  \rho^2_i  {V_i^1}^\dagger \cdots  {V_i^l}^\dagger      \triangleq    \sum_i \mathcal{E}^i (\rho^1),
\end{equation}
where $\mathcal{E}^i$ is a quantum operation that acts on the state $\rho^1$.

After getting the output of the quantum circuit $\rho^3$, the cost function can be written as
\begin{equation}
C(s) = \frac{1}{N} \sum\limits_{x=1}^N  \bra{\psi_x} \rho_x^3 \ket{\psi_x} = \frac{1}{N} \mathrm{tr}( \ket{\psi_x}   \bra{\psi_x} \rho_x^3).
\end{equation}
Since 
\begin{equation}
\begin{aligned}
\rho^3(s+\epsilon) = &\sum_i  e^{i\epsilon K_{2,i}^l} V_i^l \cdots e^{i\epsilon K_{2,i}^1} V_i^1 \mathrm{tr}_A[ e^{i\epsilon K_1^k } U^k  \cdots e^{i\epsilon K_1^1 }   \\
& U^1 \rho^0_x {U^1}^\dagger e^{-i\epsilon K_1^1} \cdots {U^k}^\dagger e^{-i\epsilon K_1^k} ( \ket{i}\bra{i}_A \otimes I_B ) ] \\
& {V_i^1}^\dagger e^{-i\epsilon K_{2,i}^1} \cdots {V_i^l}^\dagger e^{-i\epsilon K_{2,i}^l},
 \end{aligned}
\end{equation} 
we can evaluate the derivative of the cost function as
\begin{equation}
\frac{ dC}{ ds } = \lim\limits_{\epsilon > 0} \frac{C(s+\epsilon)-C(s)}{\epsilon} = \frac{1}{N} \mathrm{tr}(  \ket{\psi_x} \bra{\psi_x}  X),
\end{equation}
where
\begin{equation}
\begin{aligned}
X= & \sum_{i=1}^4 \{ [i K_{2.i}^l, V_i^l \cdots V_i^1  \rho^2_i  {V_i^1}^\dagger \cdots  {V_i^l}^\dagger] +\cdots +  \\
&   V_i^l \cdots V_i^2 [ i K_{2.i}^1, V_i^1  \rho^2_i  {V_i^1}^\dagger  ] {V_i^2}^\dagger \cdots  {V_i^l}^\dagger  +  \\
& +  \mathcal{E}^i ( [iK^k_1(s), U^k \cdots U^1 \rho^0 {U^1}^\dagger \cdots {U^k}^\dagger] ) + \cdots +  \\
& +  \mathcal{E}^i ( U^k \cdots U^2 [iK^1_1(s), U^1 \rho^0 {U^1}^\dagger] {U^2}^\dagger \cdots {U^k}^\dagger )  \} .\\
\end{aligned}
\end{equation}

To update the parameter in the network, we minimize the function
\begin{equation}
\frac{dC}{ds} - \lambda \sum_{\alpha_1,\cdots, \alpha_n} K^2_{\alpha_1,\cdots, \alpha_n} (s)^2,
\end{equation}
where $K_{\alpha_1,\cdots, \alpha_n}(s)$ is related to $K_1(s)$ by
\begin{equation}
K_1(s) = \sum   K_{\alpha_1,\cdots, \alpha_n}(s) \otimes_{k=1}^n \sigma^{\alpha_k}.
\end{equation}
We will focus on two specific cases of $K_1(s)$: 
$K_1(s)  \to  \sigma_j^\alpha$ and $K_1(s)  \to  \ket{11}_{jk} $.
The former case involves one qubit, while the latter case only involves two qubits. 
If $K_1^j(s)$ only acts on qubit $j$, we let
\begin{equation}
K_1(s) = K_1^j(s) \otimes I_{\bar{j}}.
\end{equation}
If $K_1^{j,k}(s)$ acts on qubits $j$ and $k$, we let
\begin{equation}
K_1(s) = K_1^{j,k}(s) \otimes I_{\bar{j,k}}.
\end{equation}
This ends the classical simulation of the controlled unitaries.

\section{Simulation details}
\label{appsec:detail}
This section presents additional simulation setups for the figures in Sec.~\ref{sec:simulationresult}.

To know beforehand that the global optimal loss of the DIBoM can reach 0 with a suitable tuning of its parameters, we make the following restrictions on the intrinsic unitary $V$. For Figs. \ref{fig:plotHybrid}, \ref{fig:compare}, and \ref{fig:baseline}, the intrinsic unitary $V$ is restricted to a single-qubit unitary multiplied by a generalized CZ gate, with the single-qubit unitary acting on the second qubit. For Fig.~\ref{fig:plotSG}, the intrinsic unitary $V$ is restricted to a single-qubit unitary acting on the second qubit. For Fig.~\ref{fig:plotCZ}, the intrinsic unitary $V$ is restricted to a layer of generalized CZ gates. For Fig.~\ref{fig:fixedsecond}, the intrinsic unitary $V$ is chosen in such a way that it is obtained by a single-qubit gate multiplied by a generalized CZ gate, where the generalized CZ gate of $V$ is set to be identical to the fixed second layer of the DIBoM.  
For Fig.~\ref{fig:fixedfirst}, the intrinsic unitary $V$ is also obtained by a single-qubit gate multiplied by a generalized CZ gate, with the single-qubit gate matching the first layer of the DIBoM. For Figs.~\ref{fig:product_gate}, \ref{fig:variousqubit}, \ref{fig:localcost}, and \ref{fig:globalVSlocal}, the intrinsic unitary $V$ is restricted to a product gate layer followed by a generalized CZ gate layer, with the product gate acting on all qubits of the quantum input. For Fig.~\ref{fig:ab_GCZ}, the intrinsic unitary $V$ is restricted to have the same circuit structure as the DIBoM but with different parameters. For Fig.~\ref{fig:variousLayer}, the intrinsic unitary $V$ is restricted to an alternating product of a product gate layer and a generalized CZ gate layer with a total of $L$ layers, where $L$ is the given layer number. 

Some auxiliary setups are as follows. For Fig.~\ref{fig:compare}, both the training and test samples are drawn from the same distribution, meaning that the intrinsic unitary $V$ that transforms the input into the output is identical for both sets. The number of samples in the test set is fixed at 10. 
For Fig.~\ref{fig:full_unitary}, the intrinsic unitary $V$ is randomly selected from all 2-qubit unitaries, with no additional constraints. 

\section{Simulation of local cost function}
\label{appsec:localcost}

When simulating the local cost function, there are two changes compared to simulating the global cost function. Firstly, the input and output are reversed. Secondly, the input $\ket{\phi}\bra{\phi}$ is substituted by $\ket{\phi}_i \bra{\phi} \otimes I_{\bar{i}}$.

Let us start with the first change. The ground truth unitary is $V$, hence the ground truth output is $\ket{\phi_x^L} = V\ket{\phi_x^0}$, where $\ket{\phi_x^0}$
is the quantum input. In the reverse setup, we will compare the ``model'' input $ \rho_x^0 = U^\dagger \ket{\phi_x^L}\bra{\phi_x^L} U$  with the actual input $\ket{\phi_x^0}\bra{\phi_x^0}$ in the cost function:
\begin{equation}
C_{reverse}(s) = \frac{1}{N} \sum \limits_{x=1}^N \bra{\phi_x^{0}} \rho_x^{0}(s) \ket{\phi_x^{0}}.
\end{equation}
A crucial observation is 
\begin{eqnarray}
\frac{dC_{reverse}}{ds} =\frac{i}{N} \sum\limits_x \mathrm{tr} ( \sum_{l=L}^1  M^{l}(s) K^{l}(s) ), 
 \label{eq:derivative3}
\end{eqnarray}
where $M^{l}(s)$ and $K^l(s)$ are exactly the same as the ones in Appendix~\ref{appsec:derivation}. The optimization of the reversed 
cost function $C_{reverse}$ is hence equivalent to that of the original cost function $C$, thus completing the first change.

Moving on to the second change, we note that the input $\ket{\phi_x^0}$ is a product state, which allows us to express it as 
$ \ket{\phi_x^0} = \ket{\phi_x^0}_1 \otimes \dots \otimes \ket{\phi_x^0}_n  $.
As a result, the local cost function takes the form 
\begin{equation}
C_{local}(s) = \frac{1}{nN} \sum \limits_{x=1}^N  \sum \limits_{y=1}^n \mathrm{tr}( ( \ket{\phi_x^{0}}_y\bra{\phi_x^{0}}_y \otimes I_{\bar{y}}  ) \rho_x^{0}(s) ),
\end{equation}
where $I_{\bar{y}}$ is the identity operator acting on all subsystems except the $y$-th one. 
Accordingly, the derivative of $C_{local}$ with respect to the parameter $s$ can be expressed as
\begin{equation}
\frac{dC_{local}}{ds} =\frac{i}{Nn} \sum\limits_{x}  \mathrm{tr} ( \sum_{l=1}^L  M^{l}_{local}(s) K^{l}(s) ), 
\end{equation}
where
\begin{eqnarray*}
M^{l}_{local}(s)& =&  [  \prod_{j=l}^1 U^{j}(s)( \sum\limits_{y}   \ket{\phi_x^{0}}_y\bra{\phi_x^{0}}_y \otimes I_{\bar{y}} )  \prod_{j=1}^l {U^{j}}^\dagger (s) ,    \\
 && \prod_{j=l+1}^L{U^{j}}^\dagger (s)  \rho_x^0 \prod_{j=L}^{l+1} U^{j} (s) ].
\end{eqnarray*}
The remaining procedure is identical to that of the global cost function.

\bibliographystyle{apsrev4-2}

\bibliography{BibliQTF}

\end{document}